\documentclass[12pt, draftclsnofoot, onecolumn]{IEEEtran}

\newcommand{\Rmnum}[1]{\expandafter\@slowromancap\romannumeral #1@}
\makeatother
\usepackage{cleveref}
\usepackage{graphicx}
 \usepackage{epstopdf}
\usepackage{etoolbox}
\usepackage{graphicx,cite,epsfig,amssymb,amsmath,multirow}
\usepackage{graphicx,amsmath,amssymb,amsfonts,cite}
\usepackage{subfigure}
\usepackage{subfigmat}
\usepackage{etoolbox}
\usepackage{mathrsfs}
\usepackage{float}
\usepackage{bm}
\usepackage{ulem}
\usepackage{color}
\usepackage{algorithm}
\usepackage{algorithmic}
\renewcommand{\algorithmicrequire}{\textbf{Input:}}
\renewcommand{\algorithmicensure}{\textbf{Output:}}

\makeatother

\begin{document}
\title{A Covariance-Based Hybrid Channel Feedback in FDD Massive MIMO~Systems}
\author{Shuang Qiu, David Gesbert,
{\emph{Fellow}}, {\emph{IEEE}}, Da Chen, {\emph{Member}}, {\emph{IEEE}}
and~Tao~Jiang,
{\emph{Fellow}}, {\emph{IEEE}}
    \thanks{Manuscript received February 13, 2019; revised April 24, 2019 and September 10, 2019; accepted October 6, 2019. Date of publication ...; date of current version ... . This work was supported in part by ....

Copyright (c) 2015 IEEE. Personal use of this material is permitted. However, permission to use this material for any other purposes must be obtained from the IEEE by sending a request to pubs-permissions@ieee.org.

    	S. Qiu, D. Chen and T.~Jiang are with Wuhan National Laboratory for Optoelectronics, School of Electronic Information and Communications, Huazhong
University of Science and Technology, Wuhan 430074, P. R. China. (e-mail: sqiu@hust.edu.cn, chenda@hust.edu.cn, tao.jiang@ieee.org).
    	}
    \thanks{D. Gesbert is with EURECOM, 06410 Sophia-Antipolis, France. (e-mail: David.Gesbert@eurecom.fr).}}
\maketitle

\begin{abstract}
In this paper, a novel covariance-based channel feedback mechanism is investigated for frequency division duplexing (FDD) massive multi-input multi-output (MIMO) systems.
The concept capitalizes on the notion of user statistical separability which was hinted in several prior works in the massive antenna regime but has not fully exploited so far.
We propose a hybrid statistical-instantaneous feedback mechanism
where the users are separated into two classes of feedback design based on their channel covariance.
Under the hybrid framework, each user either operates on a statistical feedback mode or quantized instantaneous channel feedback mode.
The key challenge lies in the design of a covariance-aware  classification algorithm which can handle the complex mutual interactions among all users. The classification is derived from rate bound principles and a precoding method is also devised under the mixed statistical and instantaneous feedback model.
Simulations are performed to validate our analytical results and illustrate the sum rate advantages of the proposed feedback scheme under a global feedback overhead constraint.

\end{abstract}

\section{Introduction}\label{sec_intro}
Massive multiple-input multiple-output (MIMO) is expected to be a key enabler for the next generation communication systems~\cite{RusekSPM2013,larssonCM2014}.
It has drawn considerable interest from both academia and industry for its potential energy savings and spectral efficiency gains~\cite{MarzettaNov2010, NgoApril2013}.

However, the large number of antennas brings up new challenges, one of which is the acquisition of accurate instantaneous channel state information (CSI), especially the downlink CSI.
To counteract this effect, a majority of works considered time-division duplex (TDD) mode where downlink instantaneous CSI is obtained by estimating uplink CSI via channel reciprocity~\cite{JoseTWC2011}, although the downlink CSI is not always accurate in practice due to calibration error in baseband-to-radio frequency chains~\cite{ZhangTCOMM2015}.
However, most of the current systems is based on frequency division duplex (FDD). A successful deployment of massive MIMO in FDD setting brings up a serious problem: The uplink feedback
overhead for downlink channel acquisition increases linearly with the number of antennas and quickly grows prohibitive.
In practice, the feedback channel is quantized subject to a uplink bit resource constraint\cite{LoveJSAC2008}.
This unfortunately leaves the system designer with a tough dilemma: Allow precise feedback with unbearable cost of uplink bit resources or rough quantization at the risk of high downlink interference.

To solve this well recognized problem,
a large array of strategies have been proposed to reduce \textcolor{black}{FDD-based} MIMO feedback overhead, including recent efforts tackling \textcolor{black}{FDD-based} massive MIMO specifically, such as advanced trellis-extended codebook design~\cite{ChoiTWC2015}, compressive sensing-based channel feedback reduction~\cite{KuoWCNC2012,GaoDaiTSP2015}, antenna grouping-based feedback reduction technique~\cite{LeeChoiTcomm2015}, frequency-independent parameter extraction and downlink channel reconstruction~\cite{zhangTWC2018, hanWOCC2018}, angular domain energy distribution-based channel estimation~\cite{xieTWC2018,MBKTWC2019}  etc.
Moreover, the limited channel feedback issue was also tackled by exploiting user cooperation via device-to-device communications~\cite{MaoCL2015, ChenYinTWC2017}.
The authors in~\cite{ChenYinTWC2017} adopted cooperative precoder feedback scheme among users to improve system performance.
\textcolor{black}{Furthermore, a 3D beamforming downlink transmission algorithm was proposed for FDD massive MIMO systems in \cite{XiaoTCOM2016} to greatly reduce feedback overhead with only statistical CSI.}
Quite notably, some works took advantage of spacial low-rank channel covariance exhibited in the large array regime first characterized in~\cite{AdhikaryOct.2013, YinGesbert2013} to reduce feedback information~\cite{domeneTVT2015, WagnerTIT2012, GaoDaiTSP2015, AdhikaryOct.2013, YinGesbert2013, YinGesbertOct2014}.
The key principle is that the low-rank covariance behavior stemming from finite scattering channel models can be used to project channel into a lower dimensional space with little or no loss of information  \textcolor{black}{\cite{MullerOct.2014}}.
In turn, a two-stage precoding structure was presented in~\cite{AdhikaryOct.2013} where the first-stage precoding is the key step to reduce the cost of downlink training and uplink feedback through user grouping.
Interestingly, this result prompted a series of subsequent studies on the problem of user grouping itself, such as agglomerative clustering method~\cite{SunICC2017}, density-based clustering~\cite{SunICCSN2017,GrassiEW2018} etc.

Although the above-mentioned works capitalize on the {\em low rank property} of channel covariance, they fail to exploit fully the mutual inter-covariance {\em orthogonality property} that inherently comes along with it, and not {\em for the purpose of feedback reduction}.
To further build up intuition into this issue, consider the following two examples:
First, the case of two closely spaced users whose channels undergo scattering over a limited radius around them (e.g. under the famed one ring model~\cite{AdhikaryOct.2013}).
\textcolor{black}{In this case, their signal subspaces mostly coincide. Although their instantaneous channels can be equivalently represented by their low-rank covariance's signal space projections, accurate (reduced) {\em instantaneous} CSI feedback is still required to avoid serious inter-user interference.
In the second example, these two users move far from each other and their signal subspaces become distinct.
In this case, it is well known that an interference canceling precoder can be designed based on channel covariance matrices alone\cite{AdhikaryOct.2013, YinGesbertOct2014, WangJinTSP2012}.
In other words, inter-user signal subspace orthogonality can be exploited to reduce the requirement of accurate instantaneous CSI and feedback overhead when feedback bit budget is limited.}
Interesting results were earlier reported about the impact of spatial statistics on feedback overhead~\cite{brunocorrelated}.
Elsewhere, the allocation of feedback bits was even designed as a function of transmit covariance matrix information~\cite{brunoallocation}.
However, these works exploited finely the per-user low-rank covariance properties and the inter-user orthogonality remained ignored, which leads to an identical feedback bit allocation to all the users if the users have roughly the same covariance rank and eigenvalues.

\textcolor{black}{In this paper, we highlight the fact that even when users have roughly the same covariance properties, feedback overhead can be saved by {allocating differentiated feedback bits among users with their pair-wise channel covariance orthogonality}.}
To the best of our knowledge, the pair-wise channel covariance property has not yet been exploited for feedback bit allocation.
A possible reason is the irregularity of the phenomenon: Random channel statistical behavior causes a variety of ranks to be observed in channel covariance as well as highly diverse ``degrees" of orthogonality between pairs of users, making it very difficult in practice to assign a rate-optimal amount of instantaneous CSI feedback bits to each user.
This work counteracts this issue by proposing a novel simplification strategy for feedback assignment.
Our basic concept lies in a binary version of the hybrid statistical-instantaneous feedback scheme.
\textcolor{black}{Under this feedback concept, each user is classified either as an {\em instantaneous feedback} user (labeled as class-I user) or a {\em statistical feedback} user (labeled as class-S user). More classes could be considered in principle but are fairly challenging, which are left out for further studies.}
The classification is assumed to be carried out as a preamble on the basis of statistical information alone (covariance matrices).
The challenge lies coming up with an optimal classification algorithm capable of processing the complex mutual interactions among the covariance matrices of users.

The solution of this problem is carried out in three steps. First, we articulate a precoder design capable of handling the mixed statistical-instantaneous type of
feedback information, which can be seen as a relatively straightforward extension of both the statistical signal-to-leakage-and-noise
ratio (SLNR) \cite{WangJinTSP2012} and instantaneous SLNR precoders \cite{PatchCL2012}. Second, we present a rate bound analysis predicting the rate performance under the above precoder and any user classification solution.
Finally, a sum rate bound is derived and exploited to design a suboptimal greedy classifier with good performance-complexity trade-off since the optimal classifier is computationally complex.
To observe substantial sum rate gains on a fair feedback rate basis, the classifier is designed under the {\em same} feedback
resource constraint as a conventional feedback scheme.
For ease of exposition, our results are mainly presented in a single-cell setting (interference of intra-cell nature only). The accounting of the multi-cell case is discussed in Section \ref{sec_multicell}\footnote{\textcolor{black}{Note that the idea of hybrid feedback has been presented in our previous conference paper~\cite{QiuGesbertConf2018}.
However, there are differences.
First, the full rate analysis is explicit here while only a sketch was given before.
Secondly, we improve the method to quantize the instantaneous CSI of class-I users and more accurate quantized instantaneous CSI is obtained.
Furthermore, the multi-cell scenario is considered in this paper.}}.

The rest of the paper is organized as follows. In Section~\ref{sec_sys_mod}, the system and channel models are described.
In Section~\ref{sec_precoding}, the SLNR-based precoder is proposed for both class-I and class-S users.
The system sum rate bound for single cell setting is derived based on channel covariance in Section~\ref{sec_sum_rate_analysis}.
A user classification method is elaborated under the criterion of system sum rate maximization in Section~\ref{sec_classify_alg}.
The user classification for multi-cell scenario is given in Section~\ref{sec_multicell}. The simulation results and conclusions are presented in Section~\ref{sec_simulation} and~\ref{sec_conclusion}, respectively.

$\emph{Notations:}$ 
Boldface lowercase (uppercase) letters denote column vectors (matrices).
The superscripts~$(\cdot)^H$ represents conjugate transpose and~$\mathrm{E}\left\{ \cdot \right\}$ denote expectation operation.
The notation~$\mathbb{C}^{m \times n}$ represents a set of~$m \times n$ matrices with complex entries and~$\triangleq$ is used to denote a definition.
An~$n\times n$ identity matrix is denoted as~${{\mathbf{I}}_{n}}$ and~$\mathbf{A}=\mathrm{diag}(a_1,\dots,a_l,\dots,a_M)$ denotes a diagonal matrix whose $l$-th diagonal element is~$\left[\mathbf{A}\right]_l=a_l$.
The notations~$\lfloor x \rceil $, $\lfloor x \rfloor$ and $\lceil x \rceil$ imply rounding a decimal number to its nearest, nearest lower and nearest higher integers, respectively.
The notation~$\mathbf{z}\sim \mathcal{CN}(0,\boldsymbol{\Sigma} )$ means~$\mathbf{z}$ is a complex Gaussian random vector with zero mean and covariance matrix~$\boldsymbol{\Sigma} $.
\textcolor{black}{The vector $\mathbf{u}_{\mathrm{max}} (\mathbf{A})$ denotes the eigenvector of matrix $\mathbf{A}$ corresponding to its maximum eigenvalue $\lambda_{\mathrm{max}} (\mathbf{A})$.}
In addition, we use~$ \mathcal{X}=\{x_1, \dots, x_N \}$ and~$| \mathcal{X}  |$ to denote a set and its cardinal number, respectively.

\section{System and Channel Models}\label{sec_sys_mod}


\subsection{Channel Model}
A single-cell massive MIMO system is considered where the BS is equipped with $M$ antennas and simultaneously serves~$K$ single-antenna users labeled as user set $ \mathcal{K} = \{1, \dots, K \}$.
For each user $ k \in  \mathcal{K} $, a physical channel model which describes the multiple paths propagation is exploited and given as \cite{KuoWCNC2012, {xieTWC2018}, YinGesbert2013}
\begin{equation}\label{eq 10}
{\mathbf{h}_{k}} \triangleq \frac{1}{\sqrt{P}} \sum\limits_{p = 1}^P {\gamma _{{kp}}\mathbf{a} \left({\theta _{{kp}}}\right)},
\end{equation}
where $P$ is the number of independent, identically distributed (i.i.d.) paths,~$\gamma _{{kp}}$ represents the complex gain of the~$p$-th path, $\mathbf{a} \left({\theta _{{kp}}}\right)$ is the steering vector.
For tractability, we consider a uniform linear array (ULA) and the steering vector is given~as
\begin{equation}\label{eq 1111}
\mathbf{a}({\theta }_{{kp}} ) \triangleq  \left[ {\begin{array}{*{20}{c}}
\hspace{-0.5em} 1, \hspace{-0.5em}&{{e^{ j2\pi \frac{d}{\lambda }\sin ({\theta }_{{kp}} )}}} ,\hspace{-0.7em} & \cdots ,\hspace{-0.7em} & {{e^{ j2\pi \frac{{(M - 1)d}}{\lambda }\sin ({\theta }_{{kp}} )}}}
\end{array}}\hspace{-0.5em}  \right]^T,
\end{equation}
where ${\theta }_{{kp}}$ is the random angle of arrival (AoA) corresponding to the~$p$-th path, $d$ is the antenna spacing at the BS and $\lambda$ is wavelength.
The AoAs of the~$P$ paths are assumed to be uniformly distributed over~$\left[ \overline{\theta}_k -\theta_{\Delta}/2, \overline{\theta}_k +\theta_{\Delta}/2\right]$ where~$\overline{\theta}_k \in \left[ {-\frac{\pi}{2},\frac{\pi}{2}} \right]$ is the mean AoA and $\theta_{\Delta}$ is spread AoA (SAoA).

We assume that the BS holds the statistical information of users, such as channel covariance matrix $\mathbf{\Phi}_k=\mathrm{E} \left\{ \mathbf{h}_k \mathbf{h}_k^H \right\}, k \in \mathcal{K}$. Compared to instantaneous CSI, accurate estimation of channel covariance is much easier to obtain by long-term statistics.
Furthermore, downlink channel covariance estimation for FDD systems can be estimated from uplink channel covariance matrix through certain frequency calibration processing~\cite{LiangJSAC2001}.

\subsection{Feedback Model}\label{sub_system_model}

\begin{figure}[!h]
\centering
\includegraphics[width=4.0in]{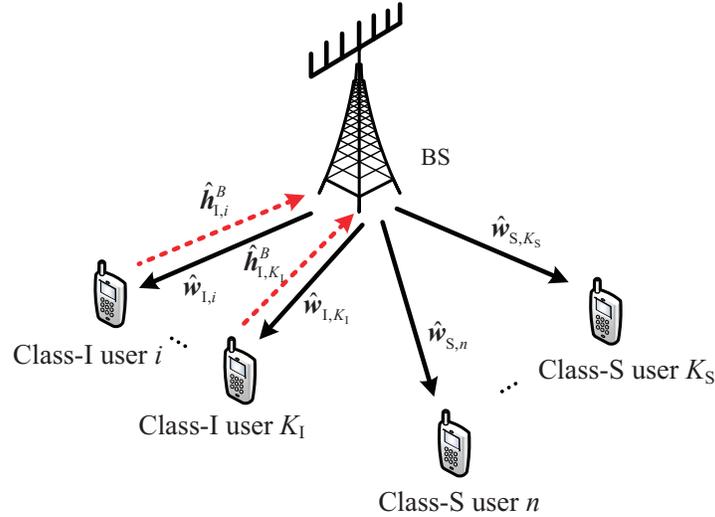}
\caption{Illustration of channel feedback and downlink data transmission under the proposed hybrid statistical-instantaneous
feedback scheme. Only class-I users feed back quantized instantaneous channel.}
\label{system_model}
\end{figure}

Under the proposed hybrid statistical-instantaneous
feedback scheme, the~$K$ users are
classified into $K_\mathrm{S}$ so-called {\em statistical feedback users} (labeled class-S users) and $K_\mathrm{I}$  {\em instantaneous feedback users} (labeled  class-I users).
The user sets are denoted as $  \mathcal{K}_\mathrm{S}=\{1, \dots, K_\mathrm{S} \}$ and~$ \mathcal{K}_\mathrm{I}=\{1, \dots, K_\mathrm{I} \}$, respectively.
Different from conventional channel feedback schemes where all the users need to feed back quantized instantaneous channel, only the class-I users feed back their quantized channel to the BS after channel quantization.
In contrast, the class-S users are assigned {\em zero} bit
towards instantaneous feedback, as shown in Fig.~\ref{system_model}.
As a result, when the total feedback bit budget is $B^{\mathrm{total}}$, each class-I user has~$B \triangleq \left\lfloor \frac{B^{\mathrm{total}}}{ K_\mathrm{I}}\right\rfloor $ bits for channel feedback.

Throughout the paper, we use subscript $(\cdot)_{\mathrm{I},i}$ and $(\cdot)_{\mathrm{S},n}$ to denote the notations for the~$i$-th class-I and~$n$-th class-S users, respectively.
The quantized channel vector $\widehat{\mathbf{h}}_{\mathrm{I},i}^{B}$ for the~$i$-th class-I user is selected based on its codebook $\mathcal{C}_{\mathrm{I},i}  \triangleq \{ \mathbf{c}_1, \dots, \mathbf{c}_{X} \}$ with $X=2^{B}$ and obtained as \cite{ChoiTWC2015}
\begin{equation}\label{1.2.1}
\widehat{\mathbf{h}}_{\mathrm{I},i}^{B}=\mathrm{arg} \mathop {{\mathop{\rm max}\nolimits} }\limits_{\mathbf{c}_u \in \mathcal{C}_{\mathrm{I},i}}  \; \left|\mathbf{h}_{\mathrm{I},i}^H \mathbf{c}_u \right|^2,
\end{equation}
where $\mathbf{h}_{\mathrm{I},i}\in \mathbb{C}^{M \times 1}$ represents the downlink instantaneous CSI\footnote{ \textcolor{black}{Downlink channel acquirement has been widely studied in FDD massive MIMO systems, such as downlink pilot signal design \cite{NohJSTSP2014}, compressed sensing-aided sparse channel estimation \cite{RaoTSP2014} etc. The overhead to obtain downlink channel can be efficiently controlled with the existing methods.
In our work, we mainly focus on uplink channel feedback issue and assume the users hold their downlink channel before conducting channel quantization and channel feedback.}}.
Thus, the BS holds the quantized channel matrix of all the class-I users as~$\widehat{\mathbf{H}}_{\mathrm{I}}^{B} \triangleq \left[ \widehat{\mathbf{h}}_{\mathrm{I},1}^{B} , \dots, \widehat{\mathbf{h}}_{\mathrm{I},K_\mathrm{I}}^{B}\right]\in \mathbb{C}^{M \times K_\mathrm{I}}$ and channel covariance matrix ~$\mathbf{\Phi}_k, k \in \mathcal{K}$ of { all} the users for downlink data transmission.
Note that our analysis does not account for the cost related to collecting covariance information, which is left out for future studies.

\subsection{Proposed Downlink Data Transmission}

As our final goal is downlink data transmission, a first challenge is how the BS can serve class-I and class-S users simultaneously without the instantaneous CSI of class-S users while managing inter-user interference.
To handle this problem, we first characterize
the received signals ${y_{\mathrm{I},i}}$ and~${y_{\mathrm{S},n}}$ at the~$i$-th class-I user and~$n$-th class-S user as
\begin{align}
 {y_{\mathrm{I},i}} & = p_d \mathbf{h}_{\mathrm{I},i}^{H}\left(\widehat{\mathbf{W}}_{\mathrm{I}} \mathbf{x}_{\mathrm{I}} + \widehat{\mathbf{W}}_{\mathrm{S}} \mathbf{x}_{\mathrm{S}}\right)+n_{\mathrm{I},i}, \label{eq 20170428.1}\\
 {y_{\mathrm{S},n}} & =p_d \mathbf{h}_{\mathrm{S},n}^{H} \left(\widehat{\mathbf{W}}_{\mathrm{I}} \mathbf{x}_{\mathrm{I}} + \widehat{\mathbf{W}}_{\mathrm{S}} \mathbf{x}_{\mathrm{S}}\right) +n_{\mathrm{S},n},\label{eq 20170428.2}
\end{align}
where $ p_d$ is downlink transmit power to each user, ${\mathbf{h}_{\mathrm{S,}n} } \in \mathbb{C}^{M \times 1}$ represents the downlink channel vector of the $n$-th class-S user, $\mathbf{x}_{\mathrm{I}} \triangleq \left[x_{\mathrm{I},1} \dots  x_{\mathrm{I},K_\mathrm{I}} \right]^T $ and~$\mathbf{x}_{\mathrm{S}} \triangleq \left[x_{\mathrm{S},1} \dots  x_{\mathrm{S},K_\mathrm{S}} \right]^T $ are consisted of downlink data symbols \textcolor{black}{satisfying $\mathrm{E} \left\{\mathbf{x}_{\mathrm{I}} \mathbf{x}_{\mathrm{I}}^H \right\}=\mathbf{I}_{K_\mathrm{I}}$ and $\mathrm{E} \left\{\mathbf{x}_{\mathrm{S}} \mathbf{x}_{\mathrm{S}}^H \right\}=\mathbf{I}_{K_\mathrm{S}}$, respectively,} $n_{\mathrm{I},i}$ and~$n_{\mathrm{S,}n}$ denote i.i.d. additive white Gaussian noise (AWGN) with zero mean and unit variance, $\widehat{\mathbf{W}}_\mathrm{I}  \triangleq \left[\widehat{\mathbf{w}}_{\mathrm{I},1},\dots,\widehat{\mathbf{w}}_{\mathrm{I},K_\mathrm{I}}\right] \in \mathbb{C}^{M \times K_\mathrm{I}}$ and $\widehat{\mathbf{W}}_\mathrm{S}\triangleq \left[\widehat{\mathbf{w}}_{\mathrm{S},1},\dots,\widehat{\mathbf{w}}_{\mathrm{S},K_\mathrm{S}}\right]  \in \mathbb{C}^{M \times K_\mathrm{S}}$ denote the precoding matrices with $\widehat{\mathbf{w}}_{\mathrm{I},i}$ and $\widehat{\mathbf{w}}_{\mathrm{S},n}$ representing the precoding vectors for the~$i$-th class-I and the $n$-th class-S users, respectively.

The received signal~${y_{\mathrm{I},i}}$
and ${y_{\mathrm{S},n}}$ are further expressed as
\vspace{-0.4em}
\begin{align}\label{eq 5}
 {y_{\mathrm{I},i}}  =&
\underbrace {p_d \mathbf{h}_{\mathrm{I},i}^{H} {\widehat{\mathbf{w}}_{\mathrm{I},i}}{x_{\mathrm{I},i}}  }_{\textrm{Expected signal}} + \underbrace {p_d \sum\limits_{j = 1, j \ne i}^{K_{\mathrm{I}}  } {\mathbf{h}_{\mathrm{I},i}^{H} {\widehat{\mathbf{w}}_ {\mathrm{I},j}}{x_{\mathrm{I},j}}}}_{\textrm{ Interference from the other class-I users}} + \underbrace {p_d \sum\limits_{n = 1}^{K_{\mathrm{S}} } \mathbf{h}_{\mathrm{I},i}^{H} {\widehat{\mathbf{w}}_{\mathrm{S},n}}{{ {x}}_{\mathrm{S},n}}}_{\textrm{ Interference from class-S users}} +\underbrace  {n_{\mathrm{I},i}}_{\textrm{AWGN}},
\end{align}
\begin{align}\label{eq 5-2}
 {y_{\mathrm{S},n}}  =&
\underbrace {p_d \mathbf{h}_{\mathrm{S},n}^{H} {\widehat{\mathbf{w}}_{\mathrm{S},n}} {x_{\mathrm{S},n}}  }_{\textrm{Expected signal}} + \underbrace {p_d \sum\limits_{q = 1, q \ne n}^{K_{\mathrm{S}}  } {\mathbf{h}_{\mathrm{S},n}^{H} {\widehat{\mathbf{w}}_ {\mathrm{S},q}}{x_{\mathrm{S},q}}}}_ {\textrm{ Interference from the other class-S users}} + \underbrace {p_d \sum\limits_{i = 1}^{K_{\mathrm{I}} } \mathbf{h}_{\mathrm{S},n}^{H} {\widehat{\mathbf{w}}_{\mathrm{I},i}}{{ {x}}_{\mathrm{I},i}}}_{\textrm{ Interference from class-I users}} +\underbrace {n_{\mathrm{S},n}}_{\textrm{AWGN}}.
\end{align}
Thus, the signal-to-interference-plus-noise ratio (SINR)~$r_{\mathrm{I},i}$ for the~$i$-th class-I user is given as
\begin{align}
r_{\mathrm{I},i}=\frac{{|{\mathbf{h}_{\mathrm{I},i}^H}{\widehat{\mathbf{w}}_{\mathrm{I},i}}{|^2}}} {{\sum\limits_{{{j = 1,j \ne i}} }^{K_{\mathrm{I}} } |{\mathbf{h}_{\mathrm{I},i}^H}{\widehat{\mathbf{w}}_{\mathrm{I},j}}{|^2} + \sum\limits_{n = 1}^{K_{\mathrm{S}}}  |{\mathbf{h}_{\mathrm{I},i}^H}{\widehat{\mathbf{w}}_{\mathrm{S},n}}{|^2} +  \frac{1}{p_d} }}.\label{sinr1}
\end{align}
Following a similar approach, the SINR of the~$n$-th class-S user~$r_{\mathrm{S},n}$ can be derived as
\begin{align}
r_{\mathrm{S},n} =\frac{{|{\mathbf{h}_{\mathrm{S},n}^H}{\widehat{\mathbf{w}}_{\mathrm{S},n}}{|^2}}} {{\sum\limits_{{{q = 1,q \ne n}} }^{K_{\mathrm{S}} } |{\mathbf{h}_{\mathrm{S},n}^H}{\widehat{\mathbf{w}}_{\mathrm{S},q}}{|^2} + \sum\limits_{i = 1}^{K_{\mathrm{I}}}  |{\mathbf{h}_{\mathrm{S},n}^H}{\widehat{\mathbf{w}}_{\mathrm{I},i}}{|^2} +  \frac{1}{p_d} }}.\label{sinr2}
\end{align}
Based on the assumption of block fading channel model, the downlink ergodic achievable rate of the~$i$-th class-I and the~$n$-th class-S users are obtained as
\begin{align}
{R_{\mathrm{I},i}} & =  \mathrm{E}\left\{ {{{  \log }_2} \left( {1 +r_{\mathrm{I},i} } \right)} \right\},\label{rate1.2.2}\\
{R_{\mathrm{S},n}} & =  \mathrm{E}\left\{ {{{  \log }_2} \left( {1 +r_{\mathrm{S},n} } \right)} \right\},\label{rate1.2.2}
\end{align}
respectively.
The system sum rate under feedback bit constraint $B^{\mathrm{total}}$ is
\begin{equation}\label{1.2.4}
R_{\mathrm{sum}}\left( {\mathcal{K} _\mathrm{I}} ,  {\mathcal{K} _\mathrm{S}} , B^{\mathrm{total}}\right) = \sum_{i=1}^{K_{\mathrm{I}}}{R_{\mathrm{I},i}} + \sum_{n=1}^{K_{\mathrm{S}}} {R_{\mathrm{S},n}}.
\end{equation}

Clearly, the system sum rate is highly influenced by the user classification solutions and the accuracy of class-I users' quantized channel.
Hence, the challenge behind this approach is to find the optimal classifier capable of leveraging the complex mutual interactions among users's channel statistics.
To solve this problem, we propose a precoder design in the next section to handle the mixed statistical-instantaneous feedback information and then derive a sum rate bound to evaluate the performance of different user classification solutions.
\section{ SLNR-Based Downlink Precoder Design}\label{sec_precoding}

Precoding methods with mixed utilization of statistical and instantaneous CSI have been studied in~\cite{QiuTVT17}, where the interference between two user classes is canceled by removing the common channel existing in overlapping subspaces.
To minimize the channel loss, a SLNR-based downlink precoder is designed in this paper with the mixed statistical-instantaneous feedback information introduced in Section \ref{sec_sys_mod}.
The motivation of using SLNR-based precoder is twofold.
First, the leakage-based criterion leads to a decoupled optimization problem and gives
an analytical closed-form precoding solution which is critical to derive the rate bounds needed for the user classification algorithm \cite{SadekTWC2007}.
Secondly, SLNR-based precoder takes the Gaussian noise into consideration and has been illustrated to achieve identical performance to minimum mean square error precoder~\cite{PatchCL2012}.

With the coexistence of class-I and class-S users, the SLNR expressions of the~$i$-th class-I and the~$n$-th class-S users~are
\begin{align}\label{3.1.1}
\Gamma_{\mathrm{I},i}=\frac{ |\mathbf{h}_{\mathrm{I},i}^H \widehat{\mathbf{w}}_{\mathrm{I},i} |^2 }{ \sum\limits_{j=1,j\neq i}^{K_\mathrm{I}} | \mathbf{h}_{\mathrm{I},j}^H \widehat{\mathbf{w}}_{\mathrm{I},i}|^2 + \sum\limits_{n=1}^{{K_\mathrm{S}}}|\mathbf{h}_{\mathrm{S},n}^H \widehat{\mathbf{w}}_{\mathrm{I},i} |^2 +\frac{1}{p_d }  } ,
\end{align}
\begin{align}\label{3.1.1}
\Gamma_{\mathrm{S},n}= \frac{  |\mathbf{h}_{\mathrm{S},n}^H \widehat{\mathbf{w}}_{\mathrm{S},n} |^2}{ \sum\limits_{q=1,q\neq n}^{K_\mathrm{S}} | \mathbf{h}_{\mathrm{S},q}^H \widehat{\mathbf{w}}_{\mathrm{S},n}|^2 + \sum\limits_{i=1}^{K_\mathrm{I}}|\mathbf{h}_{\mathrm{I},i}^H \widehat{\mathbf{w}}_{\mathrm{S},n} |^2 +\frac{1}{p_d }   }   ,
\end{align}
respectively. \textcolor{black}{Since the BS only holds the statistical CSI of class-S users, we consider average SLNR $\mathrm{E} \left\{ \Gamma_{\mathrm{I},i} \right\}$ and~$\mathrm{E} \left\{ \Gamma_{\mathrm{S},n} \right\}$ to transfer the instantaneous CSI of class-S users to their statistical CSI~\cite{WangJinTSP2012,LiTVT2016}. Therefore, the average operation is only over the instantaneous CSI $\mathbf{h}_{\mathrm{S},n}, \forall n$ of class-S users.}
By using Mullen's inequality~$\mathrm{E} \left\{ \frac{X}{Y} \right\}\geq \frac{\mathrm{E} \left\{ X\right\}}{\mathrm{E} \left\{ Y \right\}}$, the lower bounds~$\mathrm{E} \left\{ \Gamma^{\mathrm{LB}}_{\mathrm{I},i} \right\}$ and~$\mathrm{E} \left\{ \Gamma^{\mathrm{LB}}_{\mathrm{S},n} \right\}$ of the average SLNR are obtained with channel covariance~as \cite{DavidTse}
\vspace{-0.2em}
\begin{align}\label{3.1.2}
 \mathrm{E} \left\{ \Gamma_{\mathrm{I},i} \right\} \geq \mathrm{E} \left\{ \Gamma^{\mathrm{{LB}}}_{\mathrm{I},i} \right\} = \frac{  \widehat{\mathbf{w}}^H_{\mathrm{I},i} \underline{{\mathbf{H}}}_{\mathrm{I},i} \widehat{\mathbf{w}}_{\mathrm{I},i}}{\widehat{\mathbf{w}}^H_{\mathrm{I},i} \sum\limits_{{ {j = 1, j \ne i}}}^{K_{\mathrm{I}}}\underline{\mathbf{H}}_{\mathrm{I},j}  \widehat{\mathbf{w}}_{\mathrm{I},i} +
\widehat{\mathbf{w}}^H_{\mathrm{I},i} \sum\limits_{n=1}^{K_{\mathrm{S}}} \mathbf{\Phi}_{\mathrm{S},n} \widehat{\mathbf{w}}_{\mathrm{I},i}+\frac{1}{p_d}},
\end{align}
\begin{align}\label{3.1.2}
\textcolor{black}{ \mathrm{E} \left\{ \Gamma_{\mathrm{S},n} \right\} \geq \mathrm{E} \left\{ \Gamma^{\mathrm{{LB}}}_{\mathrm{S},n} \right\}  = \frac{  \widehat{\mathbf{w}}^H_{\mathrm{S},n} \mathbf{\Phi}_{\mathrm{S},n} \widehat{\mathbf{w}}_{\mathrm{S},n}} {\widehat{\mathbf{w}}^H_{\mathrm{S},n} \sum\limits_{{ {q = 1, q\ne n}}}^{K_{\mathrm{S}}} \mathbf{\Phi}_{\mathrm{S},q} \widehat{\mathbf{w}}_{\mathrm{S},n} +
\widehat{\mathbf{w}}^H_{\mathrm{S},n} \sum\limits_{i=1}^{K_{\mathrm{I}}} \underline{\mathbf{H}}_{\mathrm{I},i} \widehat{\mathbf{w}}_{\mathrm{S},n}+\frac{1}{p_d}},}
\end{align}
respectively, where $\underline{\mathbf{H}}_{\mathrm{I},i}= \widehat{\mathbf{h}}^{B}_{\mathrm{I},i} \left(\widehat{\mathbf{h}}^{B}_{\mathrm{I},i}\right)^H$ \textcolor{black}{ and $\mathbf{\Phi}_{\mathrm{S},n}= \mathrm{E}  \left\{\mathbf{h}_{\mathrm{S},n} \mathbf{h}_{\mathrm{S},n}^H  \right\}$ denotes the channel covariance matrix of the $n$-th class-S user.}
With the goal of maximizing the lower bounds~$\mathrm{E} \left\{ \Gamma^{\mathrm{LB}}_{\mathrm{I},i} \right\} $ and~$\mathrm{E} \left\{ \Gamma^{\mathrm{LB}}_{\mathrm{S},n} \right\} $, \textcolor{black}{the closed-form precoding vectors of the $i$-th class-I and the $n$-th class-S users are obtained~as}
\begin{align}
 \widehat{\mathbf{w}}_{\mathrm{I},i}  =  \mathbf{u}_\mathrm{max}\left\{ \left(\sum\limits_{{ {j = 1, j \ne i}} }^{K_{\mathrm{I}}} \underline{\mathbf{H}}_{\mathrm{I},j}+ \sum\limits_{n=1}^{K_{\mathrm{S}}} \mathbf{\Phi}_{\mathrm{S},n} +\frac{1}{p_d} \mathbf{I}_M   \right)^{-1}  \underline{\mathbf{H}}_{\mathrm{I},i}\right\},\label{snlr1}
\end{align}
\begin{align}
& \widehat{\mathbf{w}}_{\mathrm{S},n} =    \mathbf{u}_\mathrm{max}\left\{ \left( \sum\limits_{{ {q = 1, q \ne n}} }^{K_{\mathrm{S}}}   \mathbf{\Phi}_{\mathrm{S},q}+  \sum\limits_{i=1}^{K_{\mathrm{I}}} \underline{\mathbf{H}}_{\mathrm{I},i} + \frac{1}{p_d} \mathbf{I}_M  \right)^{-1} \mathbf{\Phi}_{\mathrm{S},n}\right\},
\end{align}
\textcolor{black}{respectively, $\mathbf{u}_{\mathrm{max}} (\cdot)$ denotes the eigenvector corresponding to the maximum eigenvalue.}

It can be seen that the proposed SLNR-based precoder is an extension of
the statistical SLNR and instantaneous SLNR precoders.
When all the users are selected as class-I users, the proposed precoder becomes the classical instantaneous SLNR precoder \cite{PatchCL2012}, likewise for class-S users, the proposed precoder becomes the statistical SLNR precoder\cite{WangJinTSP2012}.

\section{{System Sum Rate Bound Analysis}}\label{sec_sum_rate_analysis}

Under the proposed feedback framework, the BS needs to classify the users in the first place.
After that, the class-I users are able to quantize their instantaneous CSI according to the assigned feedback bits and pre-defined codebooks.
In other words, the BS has no any instantaneous CSI of class-I users when it performs user classification.
Therefore, system performance prediction is necessary for the BS to evaluate different user classification solutions.

In this section, we present a rate bound derived from covariance matrices alone to predict the rate performance under the proposed SLNR-based precoder and any classification solution.
The objective behind the rate bound is less to characterize precisely the system sum rate as it is to drive the design of a classification algorithm.
In the following subsections, so-called beam domain channel and channel covariance are introduced to rewrite the actual channel and channel covariance in the form of discrete Fourier transform (DFT) matrix.
Secondly, a prediction method for quantized instantaneous channel of class-I users is presented exploiting the beam domain representation and DFT matrix.

Note that the quantized instantaneous channel prediction is one-off operation and is only used for rate bound derivation.
Once the rate bound is obtained, the BS can directly use the closed-form rate bound to evaluate system performance under any user classification.

\subsection{Beam Domain Channel and Channel Covariance}\label{subsec_beam}
Channel vectors can be equivalently presented in virtual angular domain by simply sampling at equi-spaced angular intervals at the BS side.
Then, the multipath channel vector~${\mathbf{h}_{k}}, k \in \mathcal{K},$ can be approximately rewritten as a beam domain channel and given as~\cite{SunJuneTC2015,Shen2018,Xie2017}
\begin{equation}\label{eq ch1}
\begin{split}
  {{\overline{\mathbf{h}}}_{k}} & = \sum\limits_{t = 1}^M \left[ {h}^{\mathrm{BD}}_{k}\right]_t \mathbf{a}( \varphi _t)  = \mathbf{A}  {\mathbf{h}}^{\mathrm{BD}}_{k},
\end{split}
\end{equation}
where $\mathbf{A} = \left[\mathbf{a}( \varphi _1), \dots, \mathbf{a}( \varphi _t), \dots, \mathbf{a}( \varphi _M)\right] \in \mathbb{C} ^ {M \times M}$ with~$\mathbf{a}( \varphi _t)$ representing the $t$-th virtual beam and $\varphi _t$ representing its AoA, and ${\mathbf{h}}^{\mathrm{BD}}_{k}= \left[\left[ {h}^{\mathrm{BD}}_{k}\right]_1, \dots, \left[ {h}^{\mathrm{BD}}_{k}\right]_t, \dots,\left[ {h}^{\mathrm{BD}}_{k}\right]_M\right]^T $ with $\left[ {h}^{\mathrm{BD}}_{k}\right]_t$ denoting the complex gain of the~$t$-th beam.
By considering ULA with half wavelength antenna spacing, the matrix~$\mathbf{A}$ can be approximately constructed as a DFT matrix~$\mathbf{V}$~\cite{AdhikaryOct.2013, XieApri2016}.
Set~$\varphi _t =\mathrm{arcsin}(\frac{2t}{M}-1), t=1,\dots, M,$ and the $t$-th column of matrix~$\mathbf{V}$ is given as
\begin{equation}\label{eq ch2}
\mathbf{V}\left( :,t \right) \triangleq  \frac{1}{\sqrt{M}}  \left[  {\begin{array}{*{20}{c}}
  1, &{{e^{ j \pi (\frac{2t}{M}-1)}}} , & \cdots , & {{e^{ j \pi (M-1) (\frac{2t}{M}-1)}}}
\end{array}} \right]^T.
\end{equation}
Thus, the $p$-th path of the~$k$-th user can be presented with virtual beams as
\begin{equation}\label{bd eq1}
\gamma_{kp} \mathbf{a}(\theta_{kp})=\sum\limits_{t = 1}^M \left[ \widetilde{{h}}^{\mathrm{BD}}_{kp}\right]_t \mathbf{V}\left(:,t\right),
\end{equation}
where $\left[\widetilde{{h}}^{\mathrm{BD}}_{kp}\right]_t $ denotes the gain of the $p$-th path in the $t$-th virtual beam given as
\begin{align}\label{bd eq1}
\left|\left[ \widetilde{{h}}^{\mathrm{BD}}_{kp}\right]_t \right| &= \left|\gamma_{kp}\right| \mathbf{V}\left( :,t\right) ^H  \mathbf{a}(\theta_{kp}) =\frac{\left|\gamma_{kp}\right|}{\sqrt{M}} \left|e^{j\frac{M-1}{2}\pi \beta^t_{kp}} \frac{\sin\left(\frac{M}{2}\pi \beta_{kp}^t\right)}{\sin\left(\frac{1}{2}\pi \beta_{kp}^t\right)}\right|,
\end{align}
where $\beta_{kp}^t=\sin(\theta_{kp})-\frac{2 t}{M}+1$.
Then, the beam domain gain of the $k$-th user in the $t$-th beam is given as
\begin{equation}\label{bd eq2}
\left|\left[ \widetilde{{h}}^{\mathrm{BD}}_{k}\right]_t \right| = \textcolor{black}{\frac{1}{\sqrt{P}}}\sum\limits_{p = 1}^P \left|\left[ \widetilde{{h}}^{\mathrm{BD}}_{kp}\right]_t \right|.
\end{equation}
\textcolor{black}{The beam domain channel can be approximately expressed as ${{\widetilde{\mathbf{h}}}_{k}} = \sum\limits_{t = 1}^M \left|\left[ \widetilde{{h}}^{\mathrm{BD}}_{k}\right]_t \right| \mathbf{V}( :,t) $.}
The beam domain channel covariance matrix is given as
\begin{align}\label{eq ch3}
{{\widetilde{\mathbf{\Phi}}}}_k = \mathrm{E} \left\{ { { { \widetilde{{\mathbf{h}}}}}} _{k}  { {\widetilde{\mathbf{h}}}} _{k} ^H\right\}   = \mathbf{V }  {\widetilde{\mathbf{\Phi}}}^{\mathrm{BD}}_k \mathbf{V } ^H,
\end{align}
where $ \widetilde{{\mathbf{\Phi}}}^{\mathrm{BD}}_k =\mathrm{diag} \left( \mathrm{E} \left\{ \left| [ \widetilde{{h}}^{\mathrm{BD}}_{k} ]_1 \right|^2 \right\} ,\dots,  \mathrm{E} \left\{ \left| [ \widetilde{{h}}^{\mathrm{BD}}_{k} ]_M \right|^2 \right\}  \right)$.
For the simplicity of notations, we assume the complex gain of each path satisfies~$\gamma _{{kp}} \sim \mathcal{CN} (0,1)$.
Thus, the $t$-th diagonal element of $\widetilde{{\mathbf{\Phi}}}^{\mathrm{BD}}_k$ is given as
\begin{equation}\label{bd eq4}
\left[ \widetilde{{\mathbf{\Phi}}}^{\mathrm{BD}}_k \right]_t  = \mathrm{E} \left\{ \left| [ \widetilde{{h}}^{\mathrm{BD}}_{k} ]_t \right|^2 \right\}=\frac{1}{ \textcolor{black}{MP}} \sum\limits_{p = 1}^P \left| \frac{\sin\left(\frac{M}{2}\pi \beta^t_{kp}\right)}{\sin\left(\frac{1}{2}\pi \beta^t_{kp}\right)}\right|^2.
\end{equation}
It can be seen that the beam domain channel covariance is only related to the number of paths and antennas, and the AoAs of paths which can be obtained via long-term statistics.

\subsection{Quantized Instantaneous Channel Prediction}
\label{subsec_predition}
To derive a rate bound, the BS needs to know the quantized instantaneous CSI of class-I users which influences the downlink precoder design.
However, the quantized instantaneous CSI has not been fed back before user classification operation.
One possible solution is to derive the rate~bound based on predicted instantaneous CSI.

The key idea of predictting the quantized instantaneous CSI is to find the codeword from a predefined codebook which has the largest similarity to the channel direction of one user based on its beam domain channel covariance.
The codebook size decides the number of predefined spatial directions and impacts the accuracy of quantized CSI.
The detailed prediction method is given as follows.


Since the users lie in low-dimension subspaces due to limited scatterers, the codebook design for spatially correlated channel usually takes the subspaces into account~\cite{JiangCaireTWC2015}.
Therefore, we first present approximate subspaces of users with virtual beams.
Because of the low-rank property of channel covariance, the dominant nonzero elements in~$ \widetilde{{\mathbf{\Phi}}}^{\mathrm{BD}}_{\mathrm{I},i}$ are limited and assumed to be distributed between~indices $x_{{\mathrm{I},i},\mathrm{min}}$ and~$x_{{\mathrm{I},i},\mathrm{max}}$\footnote{The parameters $x_{{\mathrm{I},i},\mathrm{min}}$ and $x_{{\mathrm{I},i},\mathrm{max}}$ are influenced by the number of BS antennas $M$ and SAoA of users~\cite{XieApri2016}.
It is difficult to determine the parameters in theoretical analysis, while they can be obtained from long-term statistics or off-line tables at the~BS.}.
Then, the dominant subspace of the $i$-th class-I user can be presented~as
\begin{align}\label{eq eq2}
 \mathcal{S}_{\mathrm{I},i} =\mathrm{Span} \left\{\mathbf{V} \left(:,x\right) ,  x_{{\mathrm{I},i},\mathrm{min}}\leq x   \leq x_{{\mathrm{I},i},\mathrm{max}}   \right\}.
\end{align}
The predefined codewords are simply considered to be isotropically distributed in subspace~$ \mathcal{S}_{\mathrm{I},i} $.
Thus, the codewords $\mathbf{c}_{{\mathrm{I},i},u} \in \mathcal{C}_{\mathrm{I},i},  u=1,\dots, X,$ is created~as
\begin{equation}\label{eq ch06}
\begin{split}
\mathbf{c}_{{\mathrm{I},i},u} & =
  \frac{1}{\sqrt{M}} \left[   {\begin{array}{*{20}{c}}
 1, &{{e^{ j \pi \eta_{\mathrm{I},i}(u)}}} , & \cdots ,& {{e^{ j \pi (M-1) \eta_{{\mathrm{I},i}}(u)}}}
\end{array}}  \right]^T,
\end{split}
\end{equation}
where $\eta_{\mathrm{I},i}(u)$ is given as
\begin{equation}\label{eq theta1}
\eta_{\mathrm{I},i}(u)=\left(\frac{2x_{{\mathrm{I},i}, \mathrm{min}}}{M}-1 \right)+ u \frac{2\left(x_{{\mathrm{I},i},\mathrm{max}}- x_{{\mathrm{I},i},\mathrm{min}} \right)}{MX}.
\end{equation}
Thus, the codebook is also presented in form of DFT vectors.
Given feedback bits $B$ (codebook size $X=2^B$) for each class-I user, the codebook and quantized channel of the $i$-th class-I user can be predicted based on its beam domain channel covariance.
The codeword index and quantized channel are respectively given~as
\begin{align}
\widetilde{u}^*_{{\mathrm{I},i}} &= \mathrm{arg}\mathop {\max }\limits_{u = 1, \ldots ,X,}  \left[ { \widetilde{{\mathbf{\Phi}}}_{\mathrm{I},i}^{\mathrm{BD}}} \right]_{\left\lfloor \frac{M}{2}\left(\eta_{\mathrm{I},i}(u)+1  \right) \right\rceil}, \label{eq feedback1-1} \\
\widehat{\mathbf{h}}^B_{\mathrm{I},i}&=\mathbf{c}_ {\mathrm{I},i,\widetilde{u}^*_{{\mathrm{I},i}} }.\label{eq feedback1-2}
\end{align}
The proof of Equation (\ref{eq feedback1-1}) and (\ref{eq feedback1-2}) is given in Appendix \ref{proof_lemma1}.
\textcolor{black}{Briefly speaking,
the selected codeword of the $i$-th class-I user should be the one closest to its strongest channel direction which can be considered as the virtual beam~$m$ with the largest beam domain channel gain~$ \left[ { \widetilde{{\mathbf{\Phi}}}_{\mathrm{I},i}^{\mathrm{BD}}} \right]_{m}$.
When the number of antennas~$M$ is infinite, there must exist a codeword identical to the virtual beam~$m$, while the number of BS antennas is limited in practice.
But power leakage happens and most of power concentrates around~$m$.
Thus, the codeword corresponding to the  virtual beam~$m$ can be selected by
$\left[ { \widetilde{{\mathbf{\Phi}}}_{\mathrm{I},i}^{\mathrm{BD}}} \right]_{\left\lfloor m  \right\rceil}$ given in equation (29) and the selected codeword is taken as the predicted channel in equation~(30).}

Although the channel quantization given in (\ref{eq feedback1-2}) is not obtained from instantaneous CSI, the predicted channel can be accurate in direction based on statistical information.
Note that the quantized channel prediction is one-off operation at the BS and is only used for rate bound~derivation. The real quantized channel used for downlink data transmission will be fed back by class-I users after user classification.

\subsection{Lower Bound Analysis of System Sum Rate}

After quantized channel prediction, the BS can forecast the system sum rate with the proposed SLNR-based precoder.
First,
the downlink SLNR-based precoding vectors for the~$i$-th class-I and the~$n$-th class-S users can be approximately obtained~as
\begin{align}
\widetilde{{\mathbf{w}}}_{\mathrm{I},i}&= \mathbf{V}(:,\widetilde{\widehat{m}}_ {\mathrm{I},i}), \label{c-slnr-1}\\
\widetilde{{\mathbf{w}}}_{\mathrm{S},n}&= \mathbf{V}(:,\widetilde{l}_ {\mathrm{S},n}^*),\label{c-slnr-2}
\end{align}
respectively, where the index $\widetilde{\widehat{m}}_{\mathrm{I},i}$ is
\begin{equation}\label{eq lemma4}
\widetilde{\widehat{ {{m}}}}_{\mathrm{I},i}=\left\lfloor x_{{\mathrm{I},i},\mathrm{min}}+ \frac{x_{{\mathrm{I},i},\mathrm{max}}- x_{{\mathrm{I},i},\mathrm{min}}}{X} \widetilde{u}^*_{\mathrm{I},i}\right\rceil ,
\end{equation}
and the index $\widetilde{l}_{\mathrm{S},n}^*$ is obtained from
\begin{align}\label{eq lemma5}
{\widetilde{ {l}}^*_{\mathrm{S},n}} &= \mathrm{arg}\mathop {\max }\limits_{l = 1, \ldots ,M,}  \left[\widetilde{ {\mathbf{\Sigma}}}_{\mathrm{S},n} \right]_{l}
\end{align}
with the $l$-th diagonal element of matrix $\widetilde{ {\mathbf{\Sigma}}}_{\mathrm{S},n}$ given as
\begin{equation}\label{eq ch22}
 \left[ \widetilde{\mathbf{\Sigma}}_{\mathrm{S},n} \right]_{l}=\frac{\left[ \widetilde{{\mathbf{\Phi}}}^{\mathrm{BD}}_{\mathrm{S},n}\right]_l}{\sum\limits_{q=1,q\neq n}^{K_{\mathrm{S}}}  \left[ \widetilde{{\mathbf{\Phi}}}^{\mathrm{BD}}_{\mathrm{S},q}\right]_l+\sum\limits_{i=1}^{K_{\mathrm{I}}} \delta (\widetilde{\widehat{ {{m}}}}_{\mathrm{I},i}-l)+\frac{1}{p_d}}.
\end{equation}
The proof of equation (\ref{c-slnr-1}) and (\ref{c-slnr-2}) is given in Appendix \ref{proof_lemma2}.


Next, with the predicted quantized channel of class-I users and the approximate SLNR-based precoding vectors, a lower bound of system sum rate can be obtained as
\begin{align}\label{eq app_sumrate}
\widetilde{{R}}^{\mathrm{LB}}_ {\mathrm{sum}} &= \sum_{i=1}^{K_{\mathrm{I}}}  \log\left(1 +  \mathrm{E }\left\{\widetilde{r}^{\mathrm{LB}}_{\mathrm{I},i} \right\} \right)  + \sum_{n=1}^{K_{\mathrm{S}}}  \log\left(1+\mathrm{E }\left\{\widetilde{r}^{\mathrm{LB}}_{\mathrm{S},n}\right\} \right),
\end{align}
\textcolor{black}{where $\mathrm{E }\left\{\widetilde{r}^{\mathrm{LB}}_{\mathrm{I},i} \right\}$ and $\mathrm{E }\left\{\widetilde{r}^{\mathrm{LB}}_{\mathrm{S},n}\right\}$ denote approximate effective SINR and are respectively obtained~as}
\begin{align}\label{sinr_app1}
\mathrm{E }\left\{\widetilde{r}^{\mathrm{LB}}_{\mathrm{I},i} \right\}  =\frac{\left[ \widetilde{{\mathbf{\Phi}}} ^{\mathrm{BD}} _{\mathrm{I},i}\right]_{\widetilde{\widehat{m}} _{\mathrm{I},i}}  }{\sum\limits_{{ {j=1,j\neq i}} }^{K_{\mathrm{I}} }\left[  \widetilde{{\mathbf{\Phi}}} ^{\mathrm{BD}} _{\mathrm{I},i}\right]_{\widetilde{\widehat{m}}_ {\mathrm{I},j}}    +   \sum\limits_{n = 1}^{K_{\mathrm{S}}} \left[ \widetilde{{\mathbf{\Phi}}}^{\mathrm{BD}} _{\mathrm{I},i}\right]_{_{\widetilde{l}^*_ {\mathrm{S},n}} }  +\frac{1}{p_d}  },
\end{align}
\begin{align}\label{sinr_app2}
\mathrm{E }\left\{\widetilde{r}^{\mathrm{LB}}_{\mathrm{S},n} \right\} = \frac{\left[  \widetilde{{\mathbf{\Phi}}} ^{\mathrm{BD}} _{\mathrm{S},n}\right]_ {\widetilde{l}^*_{\mathrm{S},n}}} { \sum\limits_{{{q = 1,q \ne n}} }^{K_{\mathrm{S}} }  \left[  \widetilde{{\mathbf{\Phi}}} ^{\mathrm{BD}} _{\mathrm{S},n}\right]_ {\widetilde{l}^*_{\mathrm{S},q}} +   \sum\limits_{i = 1}^{K_{\mathrm{I}}} \left[  \widetilde{{\mathbf{\Phi}}} ^{\mathrm{BD}} _{\mathrm{S},n}\right]_ {\widetilde{\widehat{m}}_{\mathrm{I},i}} + \frac{1}{p_d} }.
\end{align}
The proof of Equation (\ref{eq app_sumrate}) is given in Appendix \ref{proof_lemma3}.

Note that the rate bound is computed based on channel statistics alone
and can be directly used to predict system performance under any user classification.
A greedy user classification algorithm is presented in the next section.

\section{Greedy User Classification}
\label{sec_classify_alg}

The optimal classifier for the proposed feedback scheme to maximize the system sum rate is computationally complex.
Therefore, the rate bound obtained in Section \ref{sec_sum_rate_analysis} is exploited to obtain a suboptimal greedy classifier with good performance-complexity trade-off.
The user classification problem can be formulated as
\begin{subequations} \label{p2}
\begin{align}
\hspace{-1em}  {\mathcal{K}^{\mathrm{sub}}_\mathrm{I}} , {\mathcal{K}^{\mathrm{sub}}_\mathrm{S}} & =  \arg \max  {\widetilde{R}^{\mathrm{LB}}_{\mathrm{sum}}}\left( {\mathcal{K} _\mathrm{I}} ,  {\mathcal{K} _\mathrm{S}}  , B^{\mathrm{total}} \right) \label{q2.1}\\
\mathrm{s. t.} \;
 \mathcal{K}  &=  \mathcal{K}_\mathrm{I}  \cup \mathcal{K}_\mathrm{S} ,\label{q2.2} \\
K &= K_\mathrm{I}+K_\mathrm{S},\label{q2.3}\\
B & =\left\lfloor \frac{B^{\mathrm{total}}}{K_{\mathrm{I}}} \right\rfloor,\label{q2.4}
\end{align}
\end{subequations}
where the system sum rate in the objective function (\ref{q2.1}) is given in (\ref{eq ch26}), the constraints~(\ref{q2.2}) and (\ref{q2.3}) is to make sure all the $K$ users are classified and each user only belongs to one user class, and the constraint (\ref{q2.4}) indicates that class-I users share the total feedback bit budget~$B^{\mathrm{total}}$~evenly.

To find the solution for problem~(\ref{p2}), a greedy user classification algorithm is proposed in Alg.~\ref{sum_rate_greedy}.
First, we assume all the~$K$ users are class-I users and calculate the predicted sum rate~$\widetilde{R}^{\mathrm{LB},K}_{\mathrm{sum}}$ based \textcolor{black}{on (36). The superscript $K$ in $ \widetilde{R}_{\mathrm{sum}}^{\mathrm{LB},K}$ denotes the number of class-I users}.
Then, we choose one user from class-I user set who can achieve the largest~$\widetilde{R}^{\mathrm{LB},K-1}_{\mathrm{sum}}$ as a new class-S user.
Repeat this procedure until all the users have been selected as class-S users.
Finally, compare all the $K+1$ sum rate~$\widetilde{R}^{\mathrm{LB},f}_{\mathrm{sum}}, f=0,\ldots,K,$ and select the largest rate with index~$d^*$.
Thus, the optimal numbers of class-I and class-S users are~$d^*-1$ and~$K+1-d^*$, respectively.
The user set of class-S users $ {\mathcal{K}^{\mathrm{sub}}_\mathrm{S}}  $ consists of the first~$K+1-d^*$ selected class-S users and the remaining users are class-I users.

\begin{algorithm}[!t]
    \renewcommand{\algorithmicrequire}{\textbf{Input:}}
	\renewcommand{\algorithmicensure}{\textbf{Output:}}
	\caption{Greedy User Classification Algorithm}
    \label{sum_rate_greedy}
	\begin{algorithmic}[1]
        \REQUIRE 
        $ {\mathbf{\Phi}}^{\mathrm{BD}}_k, k \in  \mathcal{K}$, $B^{\mathrm{total}}$
		\ENSURE $ \mathcal{K}^{\mathrm{sub}}_\mathrm{I} $, $ \mathcal{K}^{\mathrm{sub}}_\mathrm{S} $
      			\STATE \textbf{Initialize} \\ Set $f=K$ and a vector $\widetilde{\mathbf{r}}_{\mathrm{sum}}=\emptyset$\\ The set of class-I users $ \mathcal{K}_\mathrm{I} =\{1,\dots,K  \}$\\
       The set of class-S users $ \mathcal{K}_\mathrm{S} =\emptyset$\\
       Calculate $\widetilde{R}^{\mathrm{LB},f}_{\mathrm{sum}}$ based on (\ref{eq ch26})\\
       Update $\widetilde{\mathbf{r}}_{\mathrm{sum}}= \left[\widetilde{\mathbf{r}}_{\mathrm{sum}} \; \widetilde{R}^{\mathrm{LB},K}_ {\mathrm{sum}}\right]$\\
\WHILE {$f\geq 0 $}
   \STATE Decrease $f$ by 1 and calculate $B=\left\lfloor  \frac{B^{\mathrm{total}}}{f} \right\rfloor$
   \STATE Find the user with index $n_\mathrm{S}$ as class-S user satisfying
   \vspace{-0.3em}
\begin{equation}\label{eq ch30}
  n_\mathrm{S}=\arg\  \mathop {\max } \limits_{u \in   \mathcal{K}_\mathrm{I} } \widetilde{R}^{\mathrm{LB},f}_{\mathrm{sum}} \hspace{-0.3em} \left(  \mathcal{K}_\mathrm{S}  \cup \{u\} , \mathcal{K}_\mathrm{I} \setminus \{u\}, B^{\mathrm{total}} \right) \nonumber
\end{equation}
   \STATE Update $ \mathcal{K}_\mathrm{S} $ and $ \mathcal{K}_\mathrm{I}  $ as
   \vspace{-0.5em}
\begin{equation}\label{eq alg1}
\begin{split}
 \mathcal{K}_\mathrm{S}   &= \mathcal{K}_\mathrm{S}  \cup \{ n_\mathrm{S} \} \\
 \mathcal{K}_\mathrm{I} &= \mathcal{K}_\mathrm{I} \setminus \{n_\mathrm{S}\}   \nonumber
\end{split}
\end{equation}
   \STATE Update $\widetilde{\mathbf{r}}_{\mathrm{sum}}= \left[\widetilde{\mathbf{r}}_{\mathrm{sum}} \; \widetilde{R}^{\mathrm{LB},f}_{\mathrm{sum}} \right]$\\
\ENDWHILE
\STATE Find the largest rate with index $d^*$ in vector $\widetilde{\mathbf{r}}_{\mathrm{sum}}$\\
\STATE The first $K+1-d^*$ users in $ \mathcal{K}_\mathrm{S}  $ belong to~$ \mathcal{K}^{\mathrm{sub}}_\mathrm{S} $ and~$\ \mathcal{K}^{\mathrm{sub}}_\mathrm{I} $ consists of the remaining users
\STATE \textbf{Return} $ \mathcal{K}^{\mathrm{sub}}_\mathrm{I} $, $ \mathcal{K}^{\mathrm{sub}}_\mathrm{S} $
    \end{algorithmic}
\end{algorithm}

\section{User Classification for Multi-cell Scenario}\label{sec_multicell}

Different from the single-cell setting, the multi-cell scenario will give rise to inter-cell interference, especially for the users located in the edge of cells~\cite{LiuTCOM2017}.
Thus, any user classification algorithm should consider all the users in the multi-cell network to maximize the system sum rate.
In this section, we introduce the system model, precoding design, a lower bound for the system sum rate and user classification for a multi-cell network.
Note that the principles are easily derived from the single cell setting, hence only sketches of results are presented below.

An $L$-cell massive MIMO network is considered serving~$K$ users simultaneously.
We use~$\mathcal{K}_{l}^{[\mathrm{I}]}$ and $\mathcal{K}_{l}^{[\mathrm{S}]}$ to represent the user sets of class-I and class-S users in the~$l$-th cell, respectively.
The numbers of class-I and class-S users are labeled as $\left| \mathcal{K}_{l}^{[\mathrm{I}]} \right| = K_l^{[\mathrm{I}]}$ and $\left| \mathcal{K}_{l}^{[\mathrm{S}]} \right| = K_l^{[\mathrm{S}]}$, respectively.
The total numbers of class-I and class-S users in this network are~$K^{[\mathrm{I}]}$ and~$K^{[\mathrm{S}]}$, respectively.
Moreover, the channel vector between the BS in the~$l$-th cell to user~$k$ in the~$j$-th cell is modeled~as
\begin{equation}\label{eq channel-1}
\mathbf{g}_{l,j,k}=\sqrt{\varsigma_{l,j,k}}\mathbf{h}_{l,j,k},
\end{equation}
where $\varsigma_{l,j,k}$ is large-scale fading and~$\mathbf{h}_{l,j,k}$ is fast fading given in Eq.~(\ref{eq 10}).
All the class-I users in this network share the total feedback bit budge $B^{\mathrm{total}}$ evenly and each of them can be assigned~$B=\left\lfloor \frac{B^{\mathrm{total}}}{K^{[\mathrm{I}]}}\right \rfloor$ bits for channel quantization.
No cooperation is considered among the BSs and the class-I users in the $l$-th cell only feed back their quantized channel to its own BS for downlink precoding design.
Moreover, each BS is assumed to have the statistical information of all the $K$ users in this network, including the AoAs of multipaths for each user, gain variance of each path and covariance matrices.
Then, the BSs transform the statistical information into beam domain channel covariance which is composed of DFT matrix and one diagonal matrix with gain variance of virtual beams, e.g., $\widetilde{\mathbf{\Phi}}_{l,j,k}=\mathbf{V} \widetilde{\mathbf{\Phi}}^{\mathrm{BD}}_{l,j,k} \mathbf{V}^H$.
We assume that the diagonal matrices~$\widetilde{\mathbf{\Phi}}^{\mathrm{BD}}_{l,j,k}$ held by the BSs can be exchanged or sent a central control unit to conduct user~classification.

First, by considering inter-cell interference leakage, the SLNR expressions for class-I user $i$ and class-S user $n$ in the~$l$-th cell are respectively given as
\begin{equation}\label{eq m-1}
\begin{split}
&\Gamma^{[\mathrm{I}]}_{l,i}= \frac{ \left|\left(\mathbf{h}^{[\mathrm{I}]}_{l,l,i}\right)^H \widehat{\mathbf{w}}^{[\mathrm{I}]}_{l,l,i} \right|^2 }{ \sum\limits_{b \in {{\mathcal{K}^{[\mathrm{I}]}_l}\setminus \{i \}} } \left| \left(\  \mathbf{h}^{[\mathrm{I}]}_{l,l,b}\right)^H \widehat{\mathbf{w}}^{[\mathrm{I}]}_{l,l,i} \right|^2 +  \sum\limits_{n \in {{\mathcal{K}^{[\mathrm{S}]}_l}} }  \left| \left(\mathbf{h}^{[\mathrm{S}]}_{l,l,n}\right)^H \widehat{\mathbf{w}}^{[\mathrm{I}]}_{l,l,i} \right|^2 +L^{\mathrm{inter}}_{l,i}  + \frac{1}{p_d }  } ,
\end{split}
\end{equation}

\begin{equation}\label{eq m-2}
\begin{split}
&\Gamma^{[\mathrm{S}]}_{l,n}= \frac{ \left|\left(\mathbf{h}^{[\mathrm{S}]}_{l,l,n}\right)^H \widehat{\mathbf{w}}^{[\mathrm{S}]}_{l,l,n} \right|^2 }{ \sum\limits_{q \in {{\mathcal{K}^{[\mathrm{S}]}_l}}\setminus \{ n\} } \left| \left(\  \mathbf{h}^{[\mathrm{S}]}_{l,l,q}\right)^H \widehat{\mathbf{w}}^{[\mathrm{S}]}_{l,l,n} \right|^2 +  \sum\limits_{i \in {{\mathcal{K}^{[\mathrm{I}]}_l}} }  \left| \left(\mathbf{h}^{[\mathrm{I}]}_{l,l,i}\right)^H \widehat{\mathbf{w}}^{[\mathrm{S}]}_{l,l,n} \right|^2 +L^{\mathrm{inter}}_{l,n} + \frac{1}{p_d }  } ,
\end{split}
\end{equation}
where $\widehat{\mathbf{w}}^{[\mathrm{I}]}_{l,l,i}$ and $\widehat{\mathbf{w}}^{[\mathrm{S}]}_{l,l,n}$ denote the precoding vectors, $L^{\mathrm{inter}}_{l,i}$ and $L^{\mathrm{inter}}_{l,n}$  represent the interference leakage to the users in the other cells and are given as
\begin{align}\label{eq m-3}
L^{\mathrm{inter}}_{l,i}&=\sum\limits_{j \neq l } \sum\limits_{ k \in \mathcal{K}_j}  \left| \mathbf{h}_{l,j,k}^H \widehat{\mathbf{w}}^{[\mathrm{I}]}_{l,l,i} \right|^2 , \\
L^{\mathrm{inter}}_{l,n}&=\sum\limits_{j \neq l } \sum\limits_{ k \in \mathcal{K}_j}  \left| \mathbf{h}_{l,j,k}^H \widehat{\mathbf{w}}^{[\mathrm{S}]}_{l,l,n} \right|^2 ,
\end{align}
respectively. Exploiting the same idea given in Section~\ref{sec_precoding} to maximize the lower bound of average SLNR with the quantized channel of class-I users and the channel covariance matrices of the remaining users, the precoding vectors for the users in the $l$-th cell are obtained as
\begin{equation}\label{eq m-slnr1}
\widehat{\mathbf{w}}^{[\mathrm{I}]}_{l,l,i} =  \mathbf{u}_ \mathrm{max} \left\{ \left(\sum\limits_{b \in \mathcal{K}_l^{[\mathrm{I}]}\setminus \{i\}} \overline{\mathbf{H}}_{l,l,b}^{[\mathrm{I}]}  + \sum\limits_{n \in \mathcal{K}_l^{[\mathrm{S}]}} \mathbf{\Phi}^{[\mathrm{S}]}_{l,l,n}+
 \sum\limits_{j \neq l } \sum\limits_{ k \in \mathcal{K}_j} \mathbf{\Phi}_{l,j,k} + \frac{1}{p_d}\mathbf{I} \right)^{-1}  \overline{\mathbf{H}}_{l,l,i}^{[\mathrm{I}]} \right\},
\end{equation}
\begin{equation}\label{eq m-slnr2}
\widehat{\mathbf{w}}^{[\mathrm{S}]}_{l,l,n} = \mathbf{u}_ \mathrm{max}  \left\{  \left(\sum\limits_{i \in \mathcal{K}_l^{[\mathrm{I}]}} \overline{\mathbf{H}}_{l,l,i}^{[\mathrm{I}]} + \sum\limits_{q \in \mathcal{K}_l^{[\mathrm{S}]}\setminus \{n\}} \mathbf{\Phi}^{[\mathrm{S}]}_{l,l,q}+
\sum\limits_{j \neq l } \sum\limits_{ k \in \mathcal{K}_j} \mathbf{\Phi}_{l,j,k} + \frac{1}{p_d}\mathbf{I}  \right)^{-1} \mathbf{\Phi}^{[\mathrm{S}]}_{l,l,n} \right\},
\end{equation}
where $\overline{\mathbf{H}}_{l,l,i}^{[\mathrm{I}]}=\left( {\widehat{\mathbf{h}}}_{l,l,i}^{[\mathrm{I}] } \right)^H {\widehat{\mathbf{h}}}_{l,l,i}^{[\mathrm{I}] }$.
Each user in the network suffers intra-cell and inter-cell interference (from both class-I and class-S users).
Take class-I user $i$ as an example, its SINR is
\begin{equation}\label{eq m-sinr}
r_{l,l,i}^{[\mathrm{I}]}=\frac{ \left|\left(\mathbf{h}^{[\mathrm{I}]}_{l,l,i}\right)^H \widehat{\mathbf{w}}^{[\mathrm{I}]}_{l,l,i} \right|^2 }{  \sum\limits_{b \in {{\mathcal{K}^{[\mathrm{I}]}_l}\setminus \{i \}} } \left| \left(\  \mathbf{h}^{[\mathrm{I}]}_{l,l,i}\right)^H \widehat{\mathbf{w}}^{[\mathrm{I}]}_{l,l,b} \right|^2 \hspace{-0.5em}+\hspace{-0.5em} \sum\limits_{q \in {{\mathcal{K}^{[\mathrm{S}]}_l}} }  \left| \left(\mathbf{h}^{[\mathrm{I}]}_{l,l,i}\right)^H \hspace{-0.6em} \widehat{\mathbf{w}}^{[\mathrm{S}]}_{l,l,q} \right|^2 \hspace{-0.5em} +  \sum\limits_{j \neq l } \sum\limits_{ n \in \mathcal{K}_j} \left| \left(\mathbf{h}^{[\mathrm{I}]}_{j,l,i}\right)^H \hspace{-0.6em} \widehat{\mathbf{w}}_{j,j,n} \right|^2    \hspace{-0.3em}+\hspace{-0.3em} \frac{1}{p_d }  }.
\end{equation}

Then, we intend to obtain effective SINR~$\varrho_{l,l,i}^{[\mathrm{I}]}$ and~$\varrho_{l,l,n}^{[\mathrm{S}]}$ to derive a lower bound of multi-cell sum rate.
To achieve this goal,
we first calculate the predicted quantized channel for class-I users following the similar approach of Equation (\ref{eq feedback1-1}) and (\ref{eq feedback1-2}), and denote the feedback codeword index as $\widetilde{u}^{[\mathrm{I}]}_{l,l,i}$ for the class-I user $i$ in the $l$-th cell.
Then, by exploiting the predicted channels and beam domain covariance matrices, approximate precoding vectors of the users in the $l$-th cell are obtained with the similar procedure given in Equation (\ref{c-slnr-1}) and (\ref{c-slnr-2}),  and presented as~$\widetilde{{\mathbf{w}}}_{l,l,k}= \mathbf{V}\left(:, \overline{m}_{l,l,k}\right),k \in \mathcal{K}_l$.
Due to the limited space, we omit the details to obtain $\overline{m}_{l,l,k}$ and directly present the result for a class-I user as~$\overline{m}_{l,l,k}= \left\lfloor x_{{\mathrm{I},i},\mathrm{min}}+\frac{x_ {{\mathrm{I},i},\mathrm{max}}-x_{{\mathrm{I},i}, \mathrm{min}}}{2^B} \widetilde{u}^{[\mathrm{I}]}_{l,l,i} \right\rceil $.
For a class-S user,  we have
\begin{equation}\label{eq m-5}
\overline{m}_{l,l,k}=\mathrm{arg}\mathop {\max }\limits_{x = 1, \ldots ,M}  \left[ {\widetilde{\mathbf{\Sigma}}^{[\mathrm{S}]}}_{l,l,k} \right]_{x},
\end{equation}
where ${\widetilde{\mathbf{\Sigma}}}_{l,l,k}$ is a diagonal matrix and its $x$-th element is given as
\begin{equation}\label{eq ch11}
\left[ {\widetilde{\mathbf{\Sigma}}^{[\mathrm{S}]}}_{l,l,k} \right]_{x}=\frac{\left[ \widetilde{\mathbf{\Phi}}^{\mathrm{[S],BD}}_{l,l,k}\right]_x}
{\sum\limits_{q \in \mathcal{K}_l^{[\mathrm{S}]}\setminus \{k\}} \left[ \widetilde{\mathbf{\Phi}}^{\mathrm{[S],BD}}_{l,l,q} \right]_x + \sum\limits_{i \in \mathcal{K}_l^{[\mathrm{I}]}} \delta (\overline{m}_{l,l,i}-x)+\sum\limits_{j \neq l } \sum\limits_{ k \in \mathcal{K}_j} \left[\widetilde{\mathbf{\Phi}}^{\mathrm{BD}}_{l,j,k} \right]_x+   \frac{1}{p_d}},
\end{equation}
where the superscript $(\cdot)^{\mathrm{[S],BD}}$ denotes that the user belongs to class-S users. Combining with the beam domain channel representation and the approximate precoding vectors, the effective SINR can be obtained based on (\ref{eq m-sinr})~as
\begin{align}\label{sinr_app1}
\varrho_{l,l,i}^{[\mathrm{I}]}=\frac{\left[ \widetilde{{\mathbf{\Phi}} } ^{\mathrm{[I],BD}} _{l,l,i}\right]_{\overline{m}_{l,l,i}}}   {\sum\limits_{b \in \mathcal{K}^{[\mathrm{I}]}_l \setminus \{i \}}\left[  \widetilde{{\mathbf{\Phi}}} ^{\mathrm{[I],BD}} _{l,l,i}\right]_{\overline{m}_{l,l,b}}   \hspace{-0.2em} +   \hspace{-0.2em} \sum\limits_{q \in \mathcal{K}^{[\mathrm{S}]}_l} \left[ \widetilde{{\mathbf{\Phi}}}^{\mathrm{[I],BD}} _{l,l,i}\right]_{\overline{m}_{l,l,q}} \hspace{-0.2em} +\sum\limits_{j \neq l}  \sum\limits_{ t \in \mathcal{K}_j}  \left[ \widetilde{{\mathbf{\Phi}}}^{\mathrm{[I]},\mathrm{BD}}_{j,l,i}\right]_ {\overline{m}_{j,j,t}}   +\frac{1}{p_d}  },
\end{align}
\begin{align}\label{sinr_app2}
\varrho_{l,l,n}^{[\mathrm{S}]}=\frac{\left[ \widetilde{{\mathbf{\Phi}}} ^{\mathrm{[S],BD}} _{l,l,n}\right]_{\overline{m}_{l,l,n}}}   {\sum\limits_{i \in \mathcal{K}^{[\mathrm{I}]}_l  }\left[  \widetilde{{\mathbf{\Phi}}} ^{\mathrm{[S],BD}} _{l,l,n}\right]_{\overline{m}_{l,l,i}}   \hspace{-0.2em} +   \hspace{-0.2em} \sum\limits_{q \in \mathcal{K}^{[\mathrm{S}]}_l \setminus \{n\}} \left[ \widetilde{{\mathbf{\Phi}}}^{\mathrm{[S],BD}} _{l,l,n}\right]_{\overline{m}_{l,l,q}} \hspace{-0.2em} +\sum\limits_{j \neq l}  \sum\limits_{ t \in \mathcal{K}_j}  \left[ \widetilde{{\mathbf{\Phi}}}^{\mathrm{[S]},\mathrm{BD}}_{j,l,n}\right]_ {\overline{m}_{j,j,t}}   +\frac{1}{p_d}  }.
\end{align}
Thus, the sum rate of the network is given as
\begin{equation}\label{eq ch26}
\widetilde{{R}}^{\mathrm{net,LB}}_{\mathrm{sum}} = \sum\limits_{l = 1}^L \sum\limits_{ i \in  \mathcal{K}_l^{[\mathrm{I}]} }  \log\left(1 + \varrho_{l,l,i}^{[\mathrm{I}]} \right) + \sum\limits_{l = 1}^L \sum\limits_{ n \in  \mathcal{K}_l^{[\mathrm{S}]} }  \log\left(1 + \varrho_{l,l,n}^{[\mathrm{S}]} \right).
\end{equation}
Replacing the sum rate expression as~$\widetilde{{R}}^{\mathrm{net,LB}}_{\mathrm{sum}}$ and inputting the beam domain channel covariance~$\widetilde{\mathbf{\Phi}}^{\mathrm{BD}}_{l,j,k}$ of the multi-cell network into Alg.~1, we can get the user classification result for multi-cell~network.

\section{Simulation Results}\label{sec_simulation}

In this section, the analytical rate bound and the performance of the proposed hybrid statistical-instantaneous channel feedback mechanism
are evaluated.
\textcolor{black}{For any user $k$, we set $x_{k,\mathrm{min}}=1$ and $x_{k,\mathrm{max}}=M$ for channel feedback prediction.
As a comparison, we also depict the performance of the conventional feedback scheme where all the $K$ users evenly share the feedback budget and
feed back their quantized instantaneous channel to BS. Besides, SLNR precoder is adopted for the conventional scheme~\cite{PatchCL2012}. Note that the proposed and conventional feedback schemes can work for any codebook.
In order to evaluate the advantages of the proposed feedback scheme under any codebook design, two representative codebooks with and without channel covariance are considered for the simulations, i.e., DFT-based codebook and skewed~codebook:}

\emph{1) DFT-based codebook:}
The DFT-based codebook does not take channel statistics into consideration.
When the codebook size is $X$, the $u$-th codeword $\mathbf{c}_u$ is defined as
\begin{equation}\label{eq ch6}
\mathbf{c}_u \triangleq   \frac{1}{\sqrt{M}} \left[ {\begin{array}{*{20}{c}}
\hspace{-0.3em}1,\hspace{-0.6em} &{{e^{ j \pi (\frac{2u}{X}-1)}}} ,\hspace{-0.3em} & \cdots ,\hspace{-0.6em}& {{e^{ j \pi (M-1) (\frac{2u}{X}-1)}}}
\end{array}}  \right]^T.
\end{equation}

\emph{2) Skewed codebook:} For class-I user $i$ , the codebook with size $X$ is given as
\begin{equation}\label{eq ch666}
\mathcal{C}_{\mathrm{I},i}=\left\{ \frac{\mathbf{\Phi}^{1/2}_{\mathrm{I},i}\mathbf{f}_u}{\left\| \mathbf{\Phi}^{1/2}_ {\mathrm{I},i}\mathbf{f}_u \right\|} , u =1, \dots, X \right\},
\end{equation}
where $\mathbf{f}_u \in \mathbb{C}^{M \times 1}$ is isotropically distributed on the unit-sphere.
This codebook is more efficient for spatially correlated channel than DFT-based codebook~\cite{JiangCaireTWC2015}.

\textcolor{black}{For the multi-path channel model, we assume that user $k$'s channel is composed by $P=20$ paths. All the paths are assumed to be uniformly distributed over~$\left[ \overline{\theta}_k -\theta_{\Delta}/2, \overline{\theta}_k +\theta_{\Delta}/2\right]$ where the mean AoA $\overline{\theta}_k$ is uniformly distributed in $ \left[ {-\frac{\pi}{2},\frac{\pi}{2}} \right]$ and the SAoA is set to $\theta_\Delta = 10 ^\circ$ for all the simulations\cite{YinGesbert2013}.}

\subsection{Evaluation for Single-cell Scenario}

\begin{figure}[!h]
\centering
\includegraphics[width=4in]{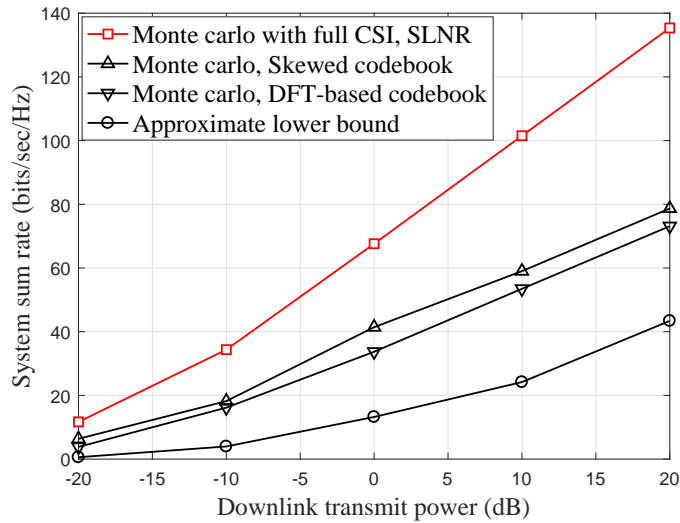}
\caption{ Performances comparison of Monte Carlo and the analytical lower bound results with $M=128$, $ K=10$ and $B^{\mathrm{total}}=40$ bits.}
\label{fig_simu_theory}
\end{figure}

Fig.~\ref{fig_simu_theory} depicts the system sum rate under the proposed feedback mechanism with Monte Carlo result and analytical lower bound derived in Section~\ref{sec_sum_rate_analysis}.
As a comparison, the Monte Carlo result of system sum rate with perfect downlink instantaneous CSI is also provided.
Fig.~\ref{fig_simu_theory} shows that the proposed channel feedback scheme achieves similar sum rate performance under the DFT-based and skewed codebooks.
Although there are only~40 feedback bits for~10 users, the proposed feedback scheme can still obtain satisfying system performance.
Moreover, although the analytical lower bound of system sum rate is derived only from channel statistics and does not rely on codebook design, it can display the change of system sum rate versus transmit power.

\begin{figure}[!h]
\centering
\includegraphics[width=4in]{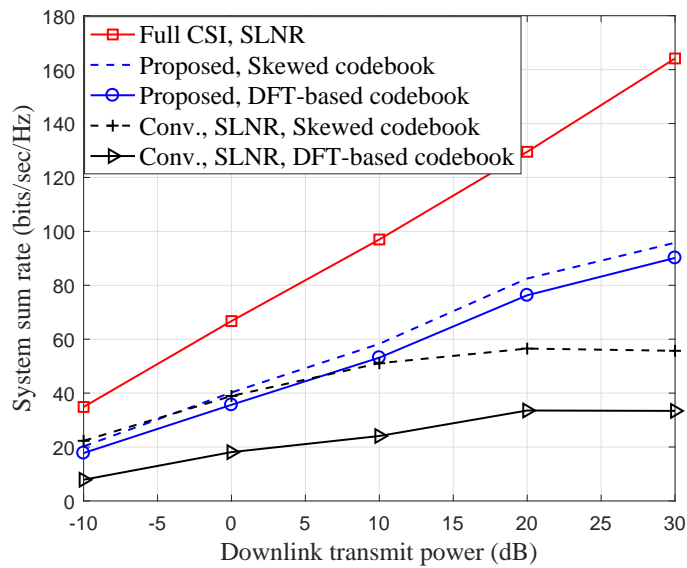}
\caption{System sum rate versus different downlink transmit power to each user with conventional and the proposed feedback schemes when $M=128$, $K=10$, $B^{\mathrm{total}}=40$ bits and $B = 4$ bits for each user under conventional~scheme.}
\label{fig_snr_sumrate}
\end{figure}

\begin{figure}[!h]
\centering
\includegraphics[width=4in]{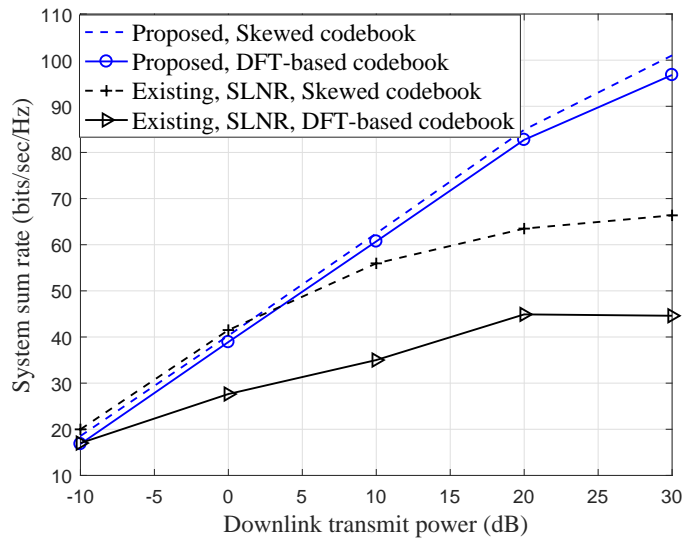}
\caption{ \textcolor{black}{Performances comparison under the proposed and the existing feedback bit allocation scheme with $M=128$, $ K=10$ and $B^{\mathrm{total}}=40$ bits.}}
\label{fig_bruno}
\end{figure}

\begin{figure}[!h]
\centering
\includegraphics[width=4in]{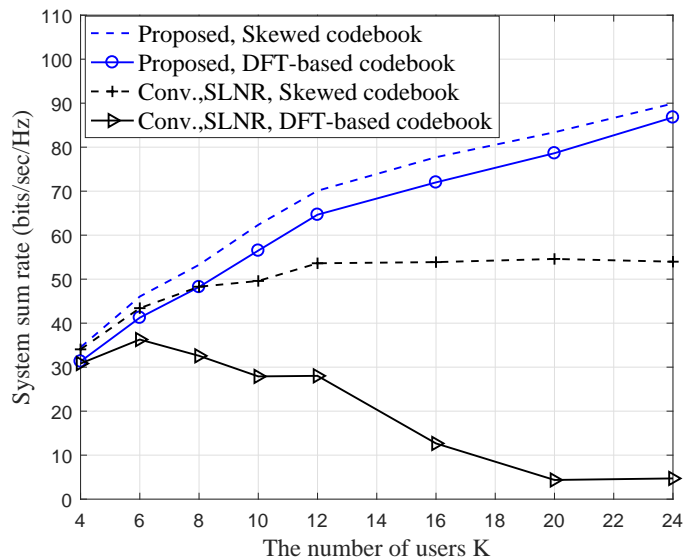}
\caption{System sum rate versus different numbers of users with conventional and the proposed feedback schemes when $M=128$, $p_d=10$ dB, $B^{\mathrm{total}}=40$ bits and $B = \left\lceil \frac{B^{\mathrm{total}}}{K}\right\rceil$ for each user under conventional~scheme.}
\label{fig_K_sumrate}
\end{figure}

\begin{figure}[!h]
\centering
\includegraphics[width=4in]{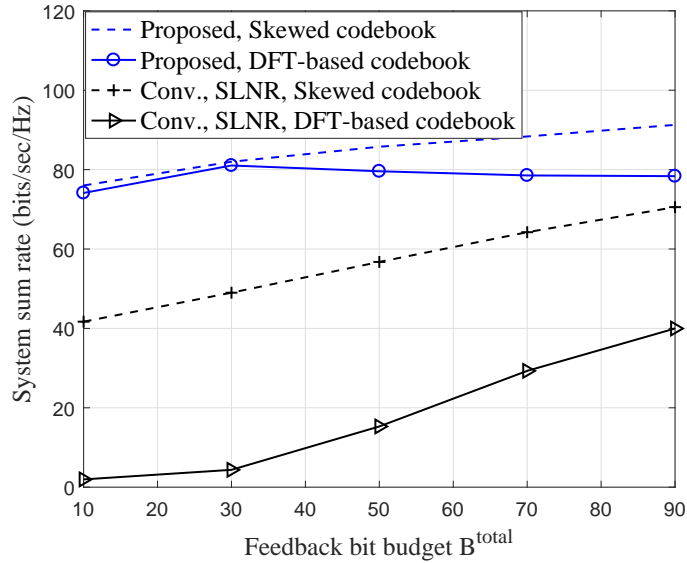}
\caption{System sum rate versus different numbers of feedback bit budget with conventional and the proposed feedback schemes when $M=128$, $K=20$, $p_d=10$ dB, $B =\left\lceil \frac{B^{\mathrm{total}}}{K} \right\rceil$ for each user under conventional~scheme.}
\label{fig_Btotal_sumrate}
\end{figure}

\begin{figure}[!h]
\centering
\includegraphics[width=4in]{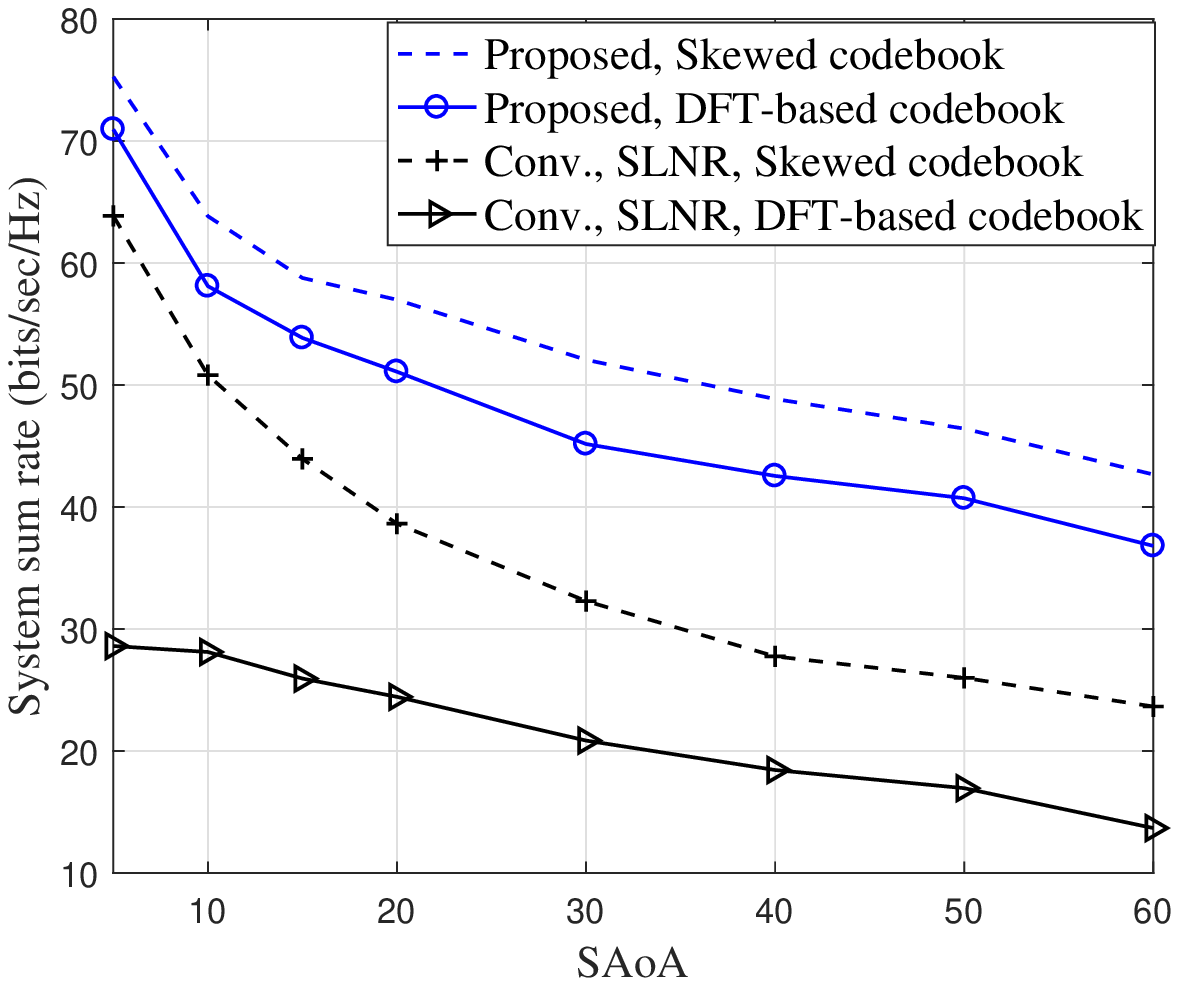}
\caption{System sum rate versus different SAoA with conventional and the proposed feedback scheme when $M=128$, $K=10$, $p_d=10$ dB, $B^{\mathrm{total}}=40$ bits and $B = 4$ bits for each user under conventional~scheme.}
\label{fig_SAoA_sumrate}
\end{figure}

The performance comparison of conventional and the proposed feedback schemes is provided in Fig.~\ref{fig_snr_sumrate} with different downlink transmit power.
Under conventional scheme, all the users share the feedback bits evenly and feed back quantized channel.
It is shown that the proposed feedback scheme outperforms the conventional one, especially when DFT-based codebook is used.
Besides, the skewed codebook achieves better performance than DFT-based codebook due to the consideration of channel statistics.
Moreover, the conventional scheme only obtains marginal performance gain in high SNR regime, while the performance of the proposed feedback scheme keeps growing with the downlink transmit power increasing.

\textcolor{black}{Fig. \ref{fig_bruno} illustrates the system sum rate under the proposed and the existing feedback bit allocation schemes \cite{brunocorrelated, brunoallocation}.
The existing works only exploited per-user low-rank covariance property to perform feedback bit allocation and the inter-user covariance orthogonality was ignored. Moreover, the feedback bit allocation of the existing works was derived under zero-forcing downlink precoder which can not handle hybrid instantaneous and statistical CSI. For fair comparison, we consider the users allocated 0 bit feedback overhead under the existing scheme as class-S users and the remaining users as class-I users. Besides, the proposed SLNR-based precoder is used for the existing feedback scheme to handle the interference among class-S and class-I users.
It can be seen from that the proposed scheme can significantly improve the system sum rate compared to the existing feedback scheme.}

Fig.~\ref{fig_K_sumrate} indicates the system sum rate versus different numbers of users under the same feedback bit budget $B^{\mathrm{total}}=40$ bits.
When only a few of users exist in the cell and each user has sufficient feedback bits (i.e., $K=4$), the BS takes every user as class-I user.
Then, the proposed feedback scheme has identical performance as conventional scheme.
Moreover, with the increasing of users, the performance of conventional scheme with DFT-based codebook badly deteriorates and the performance with skewed codebook is also restricted.
However, the performance of the proposed feedback scheme keeps growing with $K$ increasing.
When $K=20$, the system sum rate under the proposed feedback scheme is~more than 20 times larger than the conventional one with DFT-based codebook and 1.4 times larger with skewed codebook.

Fig.~\ref{fig_Btotal_sumrate} shows the system sum rate under different feedback bit budget.
The proposed scheme can always achieve
much better system performance even when the feedback bit budget is very limited, i.e., 10 bits in total for 20 users.
With the increasing of feedback bit budget, the performance of the proposed feedback scheme with skewed codebook keep rising, while the performance with DFT-based codebook slightly decreases.
When feedback bit budget is extremely large, the performances of the conventional and the proposed scheme will be~identical.

The system sum rate under different channel correlation is shown in Fig.~\ref{fig_SAoA_sumrate}.
When SAoA becomes larger, users have stronger channel correlation with the others and suffer more inter-interference.
Then, the system performance decreases under identical feedback bit budget.
However, the performance of the proposed feedback scheme outperforms the conventional scheme and achieves more stable system sum rate with the increasing SAoA.

\begin{figure}[!h]
\centering
\includegraphics[width=4in]{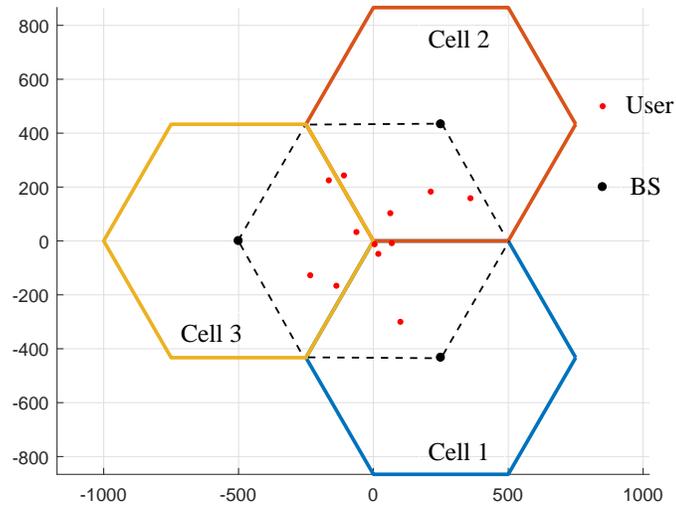}
\caption{The topology of 3-cell massive MIMO network
where only the users located in the adjacent three sectors are considered and the number of users in each cell is $K_l=4, l=1,\dots,3$.}
\label{fig_topology_multicell}
\end{figure}

\begin{figure}[!h]
\centering
\includegraphics[width=4in]{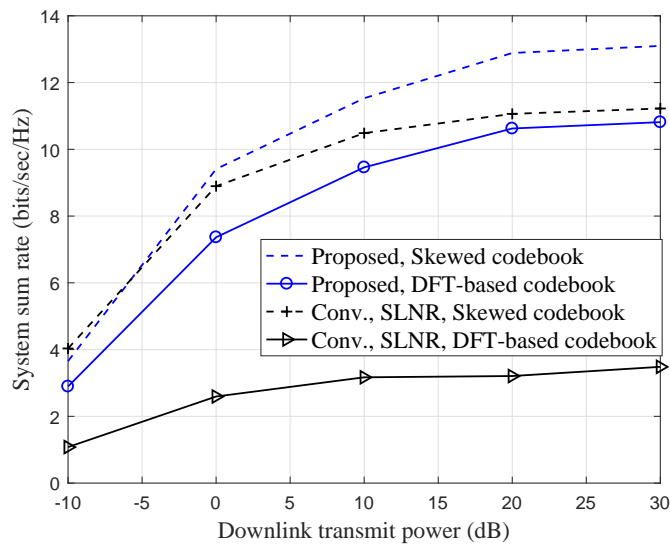}
\caption{Sum rate of multi-cell system versus transmit power with $\sigma_{\mathrm{shadow}}=8$ dB, $\nu =2.2$ for large-scale fading, $B^{\mathrm{total}}=48$ bits and $K_l=4, l=1,\dots,3$.}
\label{fig_sumrate_multicell}
\end{figure}

\subsection{Evaluation for Multi-cell Scenario}
\textcolor{black}{We consider $L=3$ cells with the radius of 500 meters and each BS is equipped with $M=128$ antennas.
We assume that the users are randomly distributed in the adjacent three sectors of the cells and no user is closer to the BSs than $r_h=100$ meters.
The topology of the system is shown in Fig. \ref{fig_topology_multicell}.
The large-scale fading is modelled as $ \varsigma_{l,j,k} = z/ (d_{l,j,k}/r_h)^\nu$, where~$z$ is a log-normal random variable with standard deviation~$\sigma_{\mathrm{shadow}}$, the variable~$d_{l,j,k}$ is the distance between user~$k$ in the~$j$-th cell to the BS in the $l$-th cell and~$\nu $ is the path loss exponent.
For fair comparison, the inter-cell interference suppression is also considered in conventional SLNR precoding scheme which is taken as a special case of~(\ref{eq m-slnr1}) and~(\ref{eq m-slnr2}) with $\mathcal{K}_l^{[\mathrm{S}]}=\emptyset, l=1,2,3$. Fig.~\ref{fig_sumrate_multicell} illustrates that the system sum rate of the proposed feedback scheme outperforms the conventional scheme, especially when DFT-based codebook is~used.}

\section{Conclusions}\label{sec_conclusion}
\textcolor{black}{This paper proposed a hybrid statistical-instantaneous channel feedback scheme for FDD-based massive MIMO systems by exploiting mutual inter-covariance orthogonality property.
Under this scheme, the BS only requires the quantized instantaneous CSI from part of users for downlink data transmission.
We developed a SLNR-based precoder to handle the mixed statistical and instantaneous channel feedback information.
Then, closed-form sum rate bounds were analyzed for both single-cell and multi-cell settings and were  used to design good performance-complexity trade-off user classification algorithms.
Simulations illustrated that the proposed feedback scheme significantly improves system sum rate over the conventional feedback schemes under feedback budget constraint, especially when the global feedback overhead is~deficient.}

\begin{appendices}
\section{ Proof of Equation (\ref{eq feedback1-1}) and (\ref{eq feedback1-2})}\label{proof_lemma1}
The quantized channel for the $i$-th class-I user is obtained~by
\begin{equation}\label{eq s1}
\begin{split}
\widehat{\mathbf{h}}_{\mathrm{I},i}^{B} =\mathrm{arg} \mathop {{\mathop{\rm max}\nolimits} }\limits_{\mathbf{c}_{{\mathrm{I},i},u} \in \mathcal{C}_{\mathrm{I},i}}  \; \left|\mathbf{h}_{\mathrm{I},i}^H \mathbf{c}_{{\mathrm{I},i},u}\right|^2,
\end{split}
\end{equation}
where $\mathbf{c}_{{\mathrm{I},i},u}$ is given in (\ref{eq ch06}) and $\left|\mathbf{h}_{\mathrm{I},i}^H \mathbf{c}_{\mathrm{I},i,u}\right|^2$ can be further derived with beam domain channel~as
\begin{align}\label{eq ch8}
& \left|\mathbf{h}_{\mathrm{I},i}^H \mathbf{c}_{\mathrm{I},i,u}\right|^2
= \sum\limits_{m=1}^M \left| \left[ {h}^{\mathrm{BD}}_{\mathrm{I},i} \right]_m \right|^2 \left|   \mathbf{V}^H\left(:,m \right)  \mathbf{c}_{\mathrm{I},i,u} \right|^2  =\frac{1}{M^2} \sum\limits_{m=1}^M \left| \left[ {h}^{\mathrm{BD}}_{\mathrm{I},i} \right]_m \right|^2 \zeta,
\end{align}
where $\zeta =  \left| e^{j \frac{M-1}{2} \pi \left(\eta_{\mathrm{I},i}(u)-\frac{2m}{M}+1\right)} \frac{\sin\left( \frac{M}{2}\pi \left( \eta_{\mathrm{I},i}(u)-\frac{2m}{M}+1 \right) \right)}{\sin\left(\frac{1}{2} \pi \left(\eta_{\mathrm{I},i}(u)-\frac{2m}{M}+1 \right) \right)}  \right|^2 $.
When the number of BS antennas satisfies $M \rightarrow \infty$, we have $\zeta \rightarrow M^2 \delta \left( \textcolor{black}{\eta}_{\mathrm{I},i}(u)-\frac{2m}{M}+1 \right)$.
Then, the expression $\left|\mathbf{h}_{\mathrm{I},i}^H \mathbf{c}_{\mathrm{I},i,u}\right|^2$ is
\begin{align}\label{eq ch9.9}
\left|\mathbf{h}_{\mathrm{I},i}^H \mathbf{c}_{\mathrm{I},i,u}\right|^2  \approx \sum\limits_{m=1}^M \left| \left[ {\widetilde{h}}^{\mathrm{BD}}_{\mathrm{I},i} \right]_m \right|^2 \delta \left( \eta_{\mathrm{I},i}(u)-\frac{2m}{M}+1  \right) .
\end{align}
Only one nonzero value exists for (\ref{eq ch9.9}) when $\eta_{\mathrm{I},i}(u)-\frac{2m}{M}+1 =0$.
\textcolor{black}{Therefore, the virtual beam of the $i$-th user corresponding to the selected codeword should be  $m_{\mathrm{I},i}=\frac{M}{2}\left(\eta_{\mathrm{I},i}(u)+1  \right)$ and the obtained value is~$\left| \left[ \widetilde{{h}}^{\mathrm{BD}}_{\mathrm{I},i} \right]_{m_{\mathrm{I},i}} \right|^2$.}
However, when $M$ is not infinite, power leakage may happen around~$m_{\mathrm{I},i}$ leading to multiple nonzero values for~(\ref{eq ch9.9}).
But most of power still concentrates around~$m_{\mathrm{I},i}$. We denote the closest beam index to~$m_{\mathrm{I},i}$ as $\widetilde{m}_{{\mathrm{I},i}}=\left\lfloor \frac{M}{2}\left(\eta_{\mathrm{I},i}(u)+1  \right) \right\rceil$.
Moreover, due to the absence of instantaneous channel gain, channel feedback is decided by channel statistics.
Thus, the value of objective function $\left| \left[ \widetilde{{h}}^{\mathrm{BD}}_{\mathrm{I},i} \right]_{\widetilde{m}_{\mathrm{I},i}} \right|^2$ corresponding to the $u$-th codebook is replaced by its variance $ \left[ { \widetilde{{\mathbf{\Phi}}}_{\mathrm{I},i}^{\mathrm{BD}}} \right]_{\widetilde{m}_{\mathrm{I},i}}$.
The codeword index for the $i$-th class-I user is
\begin{equation}\label{eq feedback1}
\widetilde{u}^*_{{\mathrm{I},i}}= \mathrm{arg}\mathop {\max }\limits_{u = 1, \ldots ,X,}  \left[ { \widetilde{{\mathbf{\Phi}}}_{\mathrm{I},i}^{\mathrm{BD}}} \right]_{\left\lfloor \frac{M}{2}\left(\eta_{\mathrm{I},i}(u)+1  \right) \right\rceil},
\end{equation}
and the quantized channel is $\widehat{\mathbf{h}}^B_{\mathrm{I},i}=\mathbf{c}_ {\mathrm{I},i,\widetilde{u}^*_{{\mathrm{I},i}} }$.

\section{Proof of Equation (\ref{c-slnr-1}) and (\ref{c-slnr-2})}\label{proof_lemma2}

For the ease of analysis for system sum rate, we first rewrite the quantized channel in the form of DFT matrix $\mathbf{V}$.
Following the similar derivation given in (\ref{eq ch8}), there exists a column vector in $\mathbf{V}$ which is identical or closest to~the predicted channel feedback~$\widehat{\mathbf{h}}^{B}_{\mathrm{I},i}$.
The index of the DFT vector satisfies
$\eta_{\mathrm{I},i}(\widetilde{u}^*_ {\mathrm{I},i})-\frac{2\widetilde{\widehat{m}}_ {\mathrm{I},i}}{M}+1=0$ and is obtained as~$\widetilde{\widehat{m}}_{\mathrm{I},i}= \left\lfloor x_{{\mathrm{I},i},\mathrm{min}}+\frac{x_ {{\mathrm{I},i},\mathrm{max}}-x_{{\mathrm{I},i}, \mathrm{min}}}{X} \widetilde{u}^*_{\mathrm{I},i}\right\rceil $.
Thus, the quantized channel~$\widehat{\mathbf{h}}^{B}_{\mathrm{I},i}$ is approximately written as~$ \widetilde{{\mathbf{h}}}^{B}_{\mathrm{I},i}=\mathbf{V} \mathbf{e}(\widetilde{\widehat{m}}_{\mathrm{I},i})$, where~$\mathbf{e}(\widetilde{\widehat{m}}_{\mathrm{I},i})$ is the $\widetilde{\widehat{m}}_{\mathrm{I},i}$-th column of an identity~matrix.

By substituting $\widetilde{{\mathbf{h}}}^{B}_{\mathrm{I},i}$ into the SLNR-based precoding vectors, an approximate precoding vector for the $i$-th class-I user is obtained as
\begin{align}
&\widetilde{{\mathbf{w}}}_{\mathrm{I},i}= \mathbf{u}_\mathrm{max}\left\{ \hspace{-0.5em} \left(\sum\limits_{j=1,j\neq i}^{K_{\mathrm{I}}} \hspace{-0.5em} \mathbf{V} \mathbf{e}(\widetilde{\widehat{m}}_{\mathrm{I},j})  \mathbf{e}^H(\widetilde{\widehat{m}}_{\mathrm{I},j}) \mathbf{V}^H+\hspace{-0.5em} \sum\limits_{n=1}^{K_{\mathrm{S}}} \mathbf{V} \widetilde{{\mathbf{\Phi}}}^{\mathrm{BD}}_{\mathrm{S},n} \mathbf{V}^H +\frac{1}{p_d} \mathbf{V} \mathbf{V}^H \right)^{-1}\hspace{-1em} \mathbf{V} \mathbf{e}(\widetilde{\widehat{m}}_{\mathrm{I},i})  \mathbf{e}^H(\widetilde{\widehat{m}}_ {\mathrm{I},i})\mathbf{V}^H\right\} \nonumber \\
&=\mathbf{u}_\mathrm{max}\left\{\mathbf{V} \left(\sum\limits_{j=1,j\neq i}^{K_{\mathrm{I}}} \mathbf{e}(\widetilde{\widehat{m}}_{\mathrm{I},j})  \mathbf{e}^H(\widetilde{\widehat{m}}_{\mathrm{I},j}) + \sum\limits_{n=1}^{K_{\mathrm{S}}}  \widetilde{{\mathbf{\Phi}}}^{\mathrm{BD}}_{\mathrm{S},n}   +\frac{1}{p_d} \mathbf{I}_M \right)^{-1} \hspace{-1em} \mathbf{e}(\widetilde{\widehat{m}}_{\mathrm{I},i})  \mathbf{e}^H(\widetilde{\widehat{m}}_{\mathrm{I},i})\mathbf{V}^H \right\}=\mathbf{V} \mathbf{e}(\widetilde{\widehat{m}}_{\mathrm{I},i})\nonumber.
\end{align}
Similarly, an approximate SLNR-based precoding vector for the~$n$-th class-S user is
\begin{align}
\widetilde{{\mathbf{w}}}_{\mathrm{S},n} &= \mathbf{u}_\mathrm{max}\left\{ \left( \sum\limits_{q=1,q\neq n}^{K_{\mathrm{S}}} \mathbf{V} \widetilde{{\mathbf{\Phi}}}^{\mathrm{BD}}_{\mathrm{S},q} \mathbf{V}^H+ \sum\limits_{i=1}^{K_{\mathrm{I}}} \mathbf{V} \mathbf{e}(\widetilde{\widehat{m}}_{\mathrm{I},i})  \mathbf{e}^H(\widetilde{\widehat{m}}_{\mathrm{I},i}) \mathbf{V}^H + \frac{1}{p_d} \mathbf{V} \mathbf{V}^H \right)^{-1} \mathbf{V} \widetilde{{\mathbf{\Phi}}}^{\mathrm{BD}}_{\mathrm{S},n} \mathbf{V}^H \right\} \nonumber \\
&= \mathbf{u}_\mathrm{max}\left\{\mathbf{V} \left( \sum\limits_{q=1,q\neq n}^{K_{\mathrm{S}}}  \widetilde{{\mathbf{\Phi}}}^{\mathrm{BD}}_{\mathrm{S},q} + \sum\limits_{i=1}^{K_{\mathrm{I}}}  \mathbf{e}(\widetilde{\widehat{m}}_{\mathrm{I},i})  \mathbf{e}^H(\widetilde{\widehat{m}}_{\mathrm{I},i}) + \frac{1}{p_d}  \mathbf{I}_M \right)^{-1} \widetilde{{\mathbf{\Phi}}}^{\mathrm{BD}}_{\mathrm{S},n} \mathbf{V}^H\right\} \nonumber \\& =\mathbf{u}_\mathrm{max}\left\{\mathbf{V} \widetilde{\mathbf{\Sigma}}_{\mathrm{S},n} \mathbf{V}^H\right\} \nonumber,
\end{align}
where
$\widetilde{\mathbf{\Sigma}}_{\mathrm{S},n}=\left( \sum\limits_{{ {q=1,q\neq n}}}^{K_{\mathrm{S}}} \hspace{-0.5em} \widetilde{{\mathbf{\Phi}}}^{\mathrm{BD}}_{\mathrm{S},q} +\hspace{-0.3em} \sum\limits_{i=1}^{K_{\mathrm{I}}}  \mathbf{E}(\widetilde{\widehat{m}}_{\mathrm{I},i}) + \frac{1}{p_d}  \mathbf{I}_M \right)^{-1} \hspace{-1em} \widetilde{{\mathbf{\Phi}}}^{\mathrm{BD}}_{\mathrm{S},n}$
with $\mathbf{E}(\widetilde{\widehat{m}}_{\mathrm{I},i})=\mathbf{e}(\widetilde{\widehat{m}}_{\mathrm{I},i})  \mathbf{e}^H(\widetilde{\widehat{m}}_ {\mathrm{I},i})$. The~$l$-th diagonal element is
\begin{equation}\label{eq ch11}
  \left[\widetilde{ \mathbf{\Sigma}}_{\mathrm{S},n} \right]_{l}=\frac{\left[ \widetilde{{\mathbf{\Phi}}}^{\mathrm{BD}}_{\mathrm{S},n}\right]_l}{\sum\limits_{q=1,q\neq n}^{K_{\mathrm{S}}}  \left[ \widetilde{{\mathbf{\Phi}}}^{\mathrm{BD}}_{\mathrm{S},q}\right]_l+\sum\limits_{i=1}^{K_{\mathrm{I}}} \delta (\widetilde{\widehat{m}}_{\mathrm{I},i}-l)+\frac{1}{p_d}}.
\end{equation}
The vector in matrix $\mathbf{V}$ corresponding to the largest~$ \left[ \widetilde{\mathbf{\Sigma}}_{\mathrm{S},n} \right]_{l}$ is selected as $\widehat{\mathbf{w}} _{\mathrm{S},n}$ and the index of the largest diagonal element is labeled as $\widetilde{l}_{\mathrm{S},n}^*$, such that~${ \widetilde{{l}}^*_{\mathrm{S},n}}= \mathrm{arg}\mathop {\max }\limits_{l = 1, \ldots ,M}  \left[ {\widetilde{\mathbf{\Sigma}}}_{\mathrm{S},n} \right]_{l}$.Therefore, the approximate precoding vector for the $n$-th class-S user is
$ \widetilde{{\mathbf{w}}}_{\mathrm{S},n}=\mathbf{V} \mathbf{e}(\widetilde{l}_{\mathrm{S},n}^*)= \mathbf{V}(:,\widetilde{l}_{\mathrm{S},n}^*)$.

\section{Proof of Equation (\ref{eq app_sumrate})}\label{proof_lemma3}
Combining with the beam domain  channel representation and the approximate precoding vectors, we obtain the lower bound SINR of the $i$-th class-I user as
\vspace{-0.2em}
\begin{align}\label{eq ch13}
{\widetilde{r}}^{\mathrm{LB}}_{\mathrm{I},i} &=\frac{\left|\left( \widetilde{{\mathbf{h}}}_{\mathrm{I},i}^{\mathrm{BD}}\right)^H \mathbf{e} \left(\widetilde{\widehat{m}}_{\mathrm{I},i} \right)\right|^2}{ \sum\limits_{j=1,j\neq i }^{K_{\mathrm{I}} } \left|\left( \widetilde{{\mathbf{h}}}_{\mathrm{I},i}^{\mathrm{BD}}\right)^H \hspace{-0.5em} \mathbf{e} \left(\widetilde{\widehat{m}}_{\mathrm{I},j} \right)\right|^2 \hspace{-0.4em} + \hspace{-0.3em}  \sum\limits_{n = 1}^{K_{\mathrm{S}}} \hspace{-0.1em} \left| \left( \widetilde{{\mathbf{h}}}_{\mathrm{I},i}^{\mathrm{BD}}\right)^H \hspace{-0.5em} \mathbf{e}(\widetilde{l}^*_{\mathrm{S},n})\right|^2 \hspace{-0.3em}+\frac{1}{p_d}   } \nonumber \\
&=\frac{\left|\left[ \widetilde{{\mathbf{h}}}^{\mathrm{BD}} _{\mathrm{I},i}\right]_{\widetilde{\widehat{m}}_ {\mathrm{I},i}}  \right|^2}{ \sum\limits_{{{ {j=1,j\neq i}}} }^{K_{\mathrm{I}} } \left|\left[ \widetilde{{\mathbf{h}}}^{\mathrm{BD}} _{\mathrm{I},i}\right]_{\widetilde{\widehat{m}}_{\mathrm{I},j}}  \right|^2 \hspace{-0.3em}+ \hspace{-0.3em}  \sum\limits_{n = 1}^{K_{\mathrm{S}}} \left|\left[  \widetilde{{\mathbf{h}}}^{\mathrm{BD}} _{\mathrm{I},i}\right]_{\widetilde{l}^*_{\mathrm{S},n}}  \right|^2+\frac{1}{p_d}  }.
\end{align}
Similarly, the lower bound SINR of the $n$-th class-S user is
\begin{align}\label{eq ch14}
{{\widetilde{r}}^{\mathrm{LB}}_{\mathrm{S},n}}
=\frac{\left|\left[ \widetilde{{\mathbf{h}}}^{\mathrm{BD}} _{\mathrm{S},n}\right]_ {\widetilde{l}^*_{\mathrm{S},n}}  \right|^2}{ \sum\limits_{{ {q=1,q\neq n}}}^{K_{\mathrm{S}} }  \left|\left[ \widetilde{ {\mathbf{h}}}^{\mathrm{BD}} _{\mathrm{S},n}\right]_{\widetilde{l}^*_ {\mathrm{S},q}}  \right|^2 \hspace{-0.3em}+   \sum\limits_{i = 1}^ {K_{\mathrm{I}}} \left|\left[ \widetilde{{\mathbf{h}}}^{\mathrm{BD}} _{\mathrm{S},n}\right]_{\widetilde{\widehat{m}} _{\mathrm{I},i}}  \right|^2+\frac{1}{p_d}  }.
\end{align}
\textcolor{black}{Since the BS only holds channel statistics, effective SINR is considered and an approximate effective SINR can be given as}
\begin{align}\label{eq ch19}
\mathrm{E }\left\{\widetilde{r}^{\mathrm{LB}}_{\mathrm{I},i} \right\}& =\frac{ \mathrm{E}\left\{ \left|\left[ \widetilde{{\mathbf{h}}}^{\mathrm{BD}} _{\mathrm{I},i}\right]_{\widetilde{\widehat{m}}_{\mathrm{I},i}}  \right|^2\right\} }{\sum\limits_{{ {j=1,j\neq i}} }^{K_{\mathrm{I}} } \mathrm{E}\left\{ \left| [ \widetilde{{\mathbf{h}} }^{\mathrm{BD}} _{\mathrm{I},i} ]_{\widetilde{\widehat{m}}_{\mathrm{I},j}}  \right|^2\right\} +   \sum\limits_{n = 1}^{K_{\mathrm{S}}} \mathrm{E}\left\{ \left| [ \widetilde{{\mathbf{h}}}^{\mathrm{BD}} _{\mathrm{I},i} ]_{\widetilde{l}^*_{\mathrm{S},n}}  \right|^2\right\}+\frac{1}{p_d}  } \nonumber \\
&=\frac{\left[ \widetilde{{\mathbf{\Phi}}} ^{\mathrm{BD}} _{\mathrm{I},i}\right]_{\widetilde{\widehat{m}} _{\mathrm{I},i}}  }{\sum\limits_{{ {j=1,j\neq i}} }^{K_{\mathrm{I}} } [  \widetilde{{\mathbf{\Phi}}} ^{\mathrm{BD}} _{\mathrm{I},i} ]_{\widetilde{\widehat{m}}_{\mathrm{I},j}}   +   \sum\limits_{n = 1}^{K_{\mathrm{S}}}  [ \widetilde{{\mathbf{\Phi}}}^{\mathrm{BD}} _{\mathrm{I},i} ]_{_{\widetilde{l}^*_{\mathrm{S},n}} }  +\frac{1}{p_d}  },
\end{align}
\begin{align}
\mathrm{E }\left\{\widetilde{r}^{\mathrm{LB}}_{\mathrm{S},n} \right\}& = \frac{\left[ \widetilde{ {\mathbf{\Phi}} } ^{\mathrm{BD}} _{\mathrm{S},n}\right]_ {\widetilde{l}^*_{\mathrm{S},n}}} { \sum\limits_{{{q = 1,q \ne n}} }^{K_{\mathrm{S}} }  \left[  \widetilde{{\mathbf{\Phi}}} ^{\mathrm{BD}} _{\mathrm{S},n}\right]_ {\widetilde{l}^*_{\mathrm{S},q}} \hspace{-0.2em} + \hspace{-0.2em}  \sum\limits_{i = 1}^{K_{\mathrm{I}}} \left[  \widetilde{{\mathbf{\Phi}}} ^{\mathrm{BD}} _{\mathrm{S},n}\right]_ {\widetilde{\widehat{m}}_{\mathrm{I},i}} \hspace{-0.2em} +\frac{1}{p_d} }.
\end{align}
Thus, the effective achievable sum rate is obtained as
\begin{align}\label{eq ch17}
\widetilde{{R}}^{\mathrm{LB}}_{\mathrm{sum}}  &=\hspace{-0.3em} \sum_{i=1}^{K_{\mathrm{I}}}  \log\left(1 \hspace{-0.3em} +\hspace{-0.3em} \mathrm{E }\left\{\widetilde{r}^{\mathrm{LB}}_{\mathrm{I},i} \right\} \right) +\hspace{-0.3em} \sum_{n=1}^{K_{\mathrm{S}}}  \log\left(1 \hspace{-0.3em} +\hspace{-0.3em} \mathrm{E }\left\{\widetilde{r}^{\mathrm{LB}}_{\mathrm{S},n}\right\} \right).
\end{align}

\end{appendices}

\end{document}